\documentclass[11.5pt,a4paper]{article}
\usepackage{float}
\usepackage{}
\usepackage{subfig}
\usepackage{amsmath}
\usepackage{amssymb}
\usepackage{graphicx}
\usepackage{hyperref}
\usepackage{csquotes}
\usepackage{multirow}
\usepackage[left=.45in,right=.45in,top=.6in,bottom=.6in,headheight=14.5pt]{geometry}
\usepackage{multirow}
\usepackage{pdflscape}
\usepackage[title]{appendix}
\usepackage[left=.45in,right=.45in,top=.6in,bottom=.6in,headheight=14.5pt]{geometry}
\usepackage{fmtcount} 
\usepackage{array,multirow}

\newcolumntype{C}[1]{>{\centering\arraybackslash}m{#1}}
\usepackage{epsfig}
\usepackage{float}
\usepackage{geometry}
\usepackage{amsmath}
\usepackage{cite}
\usepackage{ctable}

\newgeometry{vmargin={10mm,20mm},hmargin={19mm,19mm}}

\begin{document}

\title{Trimaximal Mixing and Extended Magic Symmetry in a Model of Neutrino Mass Matrix}

\author{ Labh Singh\thanks{sainilabh5@gmail.com}, Tapender\thanks{tapenderphy@gmail.com}, Monal Kashav\thanks{monalkashav@gmail.com} and Surender Verma\thanks{s\_7verma@hpcu.ac.in, Corresponding Author}}

\date{%
Department of Physics and Astronomical Science\\
Central University of Himachal Pradesh\\
Dharamshala, India 176215
}
\maketitle
\begin{abstract}
\noindent The trimaximal mixing scheme (TM$_2$) results in \textit{``magic"} neutrino mass matrix ($M_\nu$) which is known to accommodate neutrino oscillation data. In this paper, we propose a phenomenological ansatz for $M_\nu$ by extending the magic symmetry that leads to further reduction in the number of free parameters, thereby, increasing the predictability of the model. The neutrino mixing parameters, effective Majorana mass $m_{ee}$ and $CP$ invariants ($J_{CP}, I_1,I_2$) are found to exhibit strong correlations for TM$_2$ mixing paradigm. One of the generic feature of the model is the requirement of non-maximal $\theta_{23}$ for possible $CP$ violation measurable in neutrino oscillation experiments. The observables $m_{ee}$ and sum of neutrino masses ($\sum m_i$) have imperative implications for yet unknown neutrino mass hierarchy. For inverted hierarchy, the lower bound on $m_{ee}>0.02$ eV, predicted by the model, is found to be within the sensitivity reach of the $0\nu\beta\beta$ decay experiments. Also, cosmological bound of $0.12$ eV on $\sum m_i$, at 95\% CL, refutes inverted hierarchy implying TM$_2$ with normal hierarchy as the only viable possibility in the model. We have, also, illustrated a scenario wherein such a construction of the neutrino mass matrix can be realized using $\Delta(54)$ symmetry in the framework of Type-I+II seesaw mechanism.

\end{abstract}

\section{Introduction}
\noindent Within three neutrino oscillation paradigm, the neutrino mixing matrix can be parameterized in terms of three mixing angle ($\theta_{12}$, $\theta_{23}$, $\theta_{13}$) and one $CP$ phase $\delta$. The mixing angles have been measured by neutrino oscillation experiments with impressive precision. The long baseline experiments Deep Underground Neutrino Experiment (DUNE) and Tokai to Hyper-Kamiokande (T2HK) aim at measuring the yet unknown $CP$ phase $\delta$. Further, the mixing matrix is rendered more nondeterministic by existence of two additional $CP$ phases due to Majorana nature of neutrino. In light of the incomplete information about the mixing parameters, the phenomenological approaches play a pivotal role in elucidating the nature of neutrino mass matrix and underlying symmetry. For review of various approaches see Ref. \cite{Altarelli:2010gt} and references therein. The tri-bimaximal (TBM) mixing\cite{Harrison:2002er,Harrison:2002kp,Xing:2006xa,He:2003rm} is one such scenario extensively studied in the literature. TBM predicts maximal atmospheric mixing angle $\theta_{23}$ and vanishing reactor angle $\theta_{13}$. In light of observation of non-zero  $\theta_{13}$\cite{DoubleChooz:2011ymz,DayaBay:2012fng,DayaBay:2014fud,RENO:2012mkc,T2K:2013ppw} several extensions of TBM ansatz have been proposed to accommodate observed pattern of neutrino mixing\cite{Xing:2011at,Zhou:2011nu,Araki:2011wn,Haba:2011nv,Chao:2011sp,Zhang:2011aw,Rodejohann:2011uz,Marzocca:2011dh,Antusch:2011ic,Dev:2011bd,Ge:2011qn,Ge:2011ih,Ludl:2011vv,Joshipura:2011rr,Morisi:2011pm,BhupalDev:2011gi,deAdelhartToorop:2011nfg,Adulpravitchai:2011rq,Cao:2011cp,Araki:2011qy,Rashed:2011xe,Rashed:2011zs,Aranda:2011rt,Meloni:2011ac,King:2011ab}. One such possibility is ``Trimaximal mixing (TM)" pattern of neutrino mixing matrix\cite{Haba:2006dz,He:2006qd,Grimus:2008tt,Ishimori:2010fs,Shimizu:2011xg,He:2011gb,deMedeirosVarzielas:2012cet,Loualidi:2021qoq,Zhao:2020cjm,King:2019vhv,Novichkov:2018yse,Gautam:2018izb,Rodejohann:2017lre,Luhn:2013lkn,King:2011zj,Kumar:2010qz,Dev:2022krz,Grimus:2009xw,Albright:2008rp,Ding:2020vud,Zhao:2021efc} wherein if the second (first)  eigenvector remains same while first (second) and third (third) columns deviate from their TBM values it is called TM$_2$ (TM$_1$) mixing \textit{viz.,}
\begin{equation}\label{tm1}
U_{TM_1} =\begin{pmatrix}
\sqrt{\frac{2}{3}}& \frac{\cos\theta }{\sqrt{3}} & \frac{\sin\theta }{\sqrt{3}}\\
-\frac{1}{\sqrt{6}} & \frac{\cos\theta}{\sqrt{3}}-\frac{e^{i\phi}\sin\theta}{\sqrt{2}} &   \frac{\sin\theta}{\sqrt{3}}+\frac{e^{i\phi}\cos\theta}{\sqrt{2}} \\
 -\frac{1}{\sqrt{6}} & \frac{\cos\theta}{\sqrt{3}}+\frac{e^{i\phi}\sin\theta}{\sqrt{2}} &  \frac{\sin\theta}{\sqrt{3}}-\frac{e^{i\phi}\cos\theta}{\sqrt{2}}\\
\end{pmatrix},
\end{equation}

\begin{equation}\label{tm2}
U_{TM_2} =\begin{pmatrix}
\sqrt{\frac{2}{3}}\cos\theta & \frac{1}{\sqrt{3}} & \sqrt{\frac{2}{3}}\sin\theta \\
-\frac{\cos\theta}{\sqrt{6}}+\frac{e^{-i\phi}\sin\theta}{\sqrt{2}} &  \frac{1}{\sqrt{3}} & -\frac{\sin\theta}{\sqrt{6}}-\frac{e^{-i\phi}\cos\theta}{\sqrt{2}} \\
-\frac{\cos\theta}{\sqrt{6}}-\frac{e^{-i\phi}\sin\theta}{\sqrt{2}} &  \frac{1}{\sqrt{3}} & -\frac{\sin\theta}{\sqrt{6}}+\frac{e^{-i\phi}\cos\theta}{\sqrt{2}}\\
\end{pmatrix},
\end{equation}
where $\theta$ and $\phi$ are two free parameters.  It is well known that the neutrino mass  matrix obtained using the transformation

\begin{equation}\label{mnu}
    M_{\nu}=U^*_{TM_2} M_{\nu}^{d} U^{\dagger}_{TM_2},
\end{equation}
in the flavor basis, has ``magic symmetry", where $M_{\nu}^{d}=diag(m_1,m_2 e^{2i\alpha},m_3e^{2i\beta})$ is diagonal neutrino mass matrix with eigenvalues $m_i$ ($i=1,2,3$) and $\alpha,\beta$ are two Majorana phases. The magic symmetry means that the sum of the elements of row or column is equal. In general, magic neutrino mass matrix resulting from Eqn. (\ref{mnu}) is given by\cite{Lam:2006wy,Harrison:2004he,Friedberg:2006it} 

\begin{equation}\label{genmnu}
M_\nu =\begin{pmatrix}
a & b & c \\
b &  d & a+c-d \\
c &  a+c-d & b-c+d\\
\end{pmatrix},
\end{equation}
where $a$, $b$, $c$ and $d$ are, in general, complex parameters. It is evident from Eqn. (\ref{genmnu}) that row/column sum is equal to $a+b+c$  (\textit{``magic sum"}). 
In this work, given Eqn. (\ref{genmnu}), we explore a unique possibility for extension of the magic symmetry which decreases number of free parameters and, thus, increasing the predictability of the model while keeping the magical nature of the neutrino mass matrix intact. We write the Eqn. (\ref{genmnu}) as
\begin{equation}\label{exmnu}
M_\nu =\begin{pmatrix}
a & b & c \\
b &  a+b+c & -b \\
c &  -b & 2b+a\\
\end{pmatrix},
\end{equation}
assuming
\begin{equation}\label{cons}
d=a+b+c,
\end{equation}
with row/column sum equal to ``\textit{magic sum"}, as before. In Section 4, we have discussed the dynamical origin of this scenario based on $\Delta (54)$ symmetry.

\section{Formalism}
 We consider Eqn. (\ref{cons}) as additional constraint under the ambit of TM$_2$ mixing to ameliorate the allowed parameter space of the model. Using Eqns. (\ref{tm2}) and (\ref{mnu}), Eqn. (\ref{cons}) can be written as
\begin{equation}\label{cons1}
-4 e^{2 i \alpha} m_2+\sin ^2\theta \left(e^{2 i \beta} m_3+3 m_1 e^{2 i \phi}\right)-3\sqrt{3} e^{i \phi} \sin 2 \theta \left(m_1-e^{2 i \beta} m_3\right)+\cos ^2\theta \left(m_1+3 m_3 e^{2 i (\beta+\phi)}\right)=0.
\end{equation}
Eqn. (\ref{cons1}), yields two real equations \textit{viz.},
\begin{eqnarray}\label{e1}
\nonumber
&&-\frac{2}{3} m_2 \cos 2 \alpha +\frac{1}{6} m_3 \left(3 \cos ^2\theta  \cos 2 (\beta +\phi )+3 \sqrt{3} \sin 2 \theta  \cos (2 \beta +\phi )+\cos 2 \beta  \sin ^2\theta \right)+\\ &&
\frac{1}{6} m_1 \left(3 \sin ^2\theta  \cos 2 \phi -3 \sqrt{3} \sin 2 \theta  \cos \phi +\cos ^2\theta \right)=0,
\end{eqnarray}
and
\begin{eqnarray}\label{e2}
\nonumber
&&-\frac{2}{3} m_2 \sin 2 \alpha +\frac{1}{6} m_3 \left(3 \sqrt{3} \sin 2 \theta  \sin (2 \beta +\phi )+3 \cos ^2\theta  \sin 2 (\beta +\phi )+\sin 2 \beta  \sin ^2\theta \right)+\\
&&\frac{1}{6} m_1 \left(3 \sin ^2\theta  \sin 2 \phi -3 \sqrt{3} \sin 2 \theta  \sin \phi \right)=0,
\end{eqnarray}
which are, further, solved to obtain two mass ratios, $R_{21}\equiv\frac{m_2}{m_1}$ and $R_{31}\equiv\frac{m_3}{m_1}$ given by

\begin{eqnarray}\label{r21}
\nonumber
 R_{21}\equiv\frac{m_2}{m_1}=&&\frac{ -3 \sin 2 \beta  (\cos 4 \theta +3) \cos 2 \phi }{A}+\\ 
 &&  \frac{2 \left(\sin 2 \beta  \left(6 \sqrt{3} \sin 4 \theta  \cos \phi +49 \sin ^2 2 \theta \right)+6 \cos 2 \beta \left(2 \sqrt{3} \sin 2 \theta  \sin \phi -\cos 2 \theta  \sin 2 \phi \right)\right)}{A},
\end{eqnarray}
and
\begin{eqnarray}\label{r31}
R_{31}\equiv\frac{m_3}{m_1}=16\left(\frac{-3 \sin ^2\theta  \sin 2 (\alpha -\phi )+3 \sqrt{3} \sin 2 \theta  \sin (2 \alpha -\phi )-\sin 2 \alpha  \cos ^2\theta}{A}\right),
\end{eqnarray}
respectively, with
\begin{eqnarray}
\nonumber
&&A=16 \left(3 \sqrt{3} \sin 2 \theta  \sin (2 \alpha -2 \beta -\phi )+3 \cos ^2\theta  \sin 2 (\alpha -\beta -\phi )+\sin ^2\theta  \sin 2 (\alpha -\beta )\right),\\ \nonumber
&&m_2=\sqrt{m_1^2+\Delta m_{21}^2}, \quad m_3=\sqrt{m_1^2+\Delta m_{31}^2}.
\end{eqnarray}

\noindent The two mass-squared differences $\Delta m_{21}^{2}=m_{2}^{2}-m_{1}^{2}$ and $\left|\Delta m_{31}^{2}\right|=m_{3}^{2}-m_{1}^{2}$ alongwith mass ratios $\dfrac{m_1}{m_2}\equiv R_{21}$, $\dfrac{m_1}{m_3}\equiv R_{31}$ yield two values of neutrino mass $m_{1}$ given by
\begin{equation}\nonumber
m_{1}^{a}=\sqrt{\dfrac{\Delta m^2_{21}}{(R_{21})^2-1}},\hspace{3mm} m_{1}^{b}=\sqrt{\dfrac{|\Delta m^2_{31}|}{(R_{31})^2-1}},
\end{equation}
\noindent respectively.
The mass ratios in Eqns. (\ref{r21}) and (\ref{r31}) are functions of four parameters \textit{viz.}, $\theta$, $\phi$, $\alpha$ and $\beta$. The consistency of the formalism requires that two values of $m_1$ \textit{viz.} $m_1^a$, $m_1^b$ must be equal which can, further, be translated to the condition 
\begin{equation}\label{rnu}
    R_{\nu}\equiv\frac{\Delta m_{21}^2}{|\Delta m_{31}^2|}=\frac{R_{21}^2-1}{|R_{31}^2-1|}.
\end{equation}
Also, in term of elements of the TM$_2$ mixing matrix, the neutrino mixing angles can be written as
\begin{eqnarray}\label{mix}
\nonumber
    &&\sin^2\theta_{12}=\frac{|(U_{TM_2})_{12}|^2}{1-|(U_{TM_2})_{13}|^2}=\frac{1}{\cos 2 \theta+2},\\
    &&\sin^2\theta_{13}=|(U_{TM_2})_{13}|^2=
            \frac{2 \sin ^2\theta}{3},\\ \nonumber
    &&\sin^2\theta_{23}=\frac{|(U_{TM_2})_{23}|^2}{1-|(U_{TM_2})_{13}|^2}=
            \frac{1}{2} \left(\frac{\sqrt{3} \sin 2 \theta  \cos \phi }{\cos 2 \theta +2}+1\right).\\ \nonumber
\end{eqnarray}

\noindent The Jarlskog $CP$ invariant\cite{Jarlskog:1985ht,Bilenky:1987ty,Krastev:1988yu}, and other two invariants corresponding to Majorana phases $\alpha$ and $\beta$ are given by
\begin{eqnarray}\label{inv}
\nonumber
&&J_{CP}= \text{Im}[(U_{TM_2})_{11}(U_{TM_2})_{12}^*(U_{TM_2})_{21}^*(U_{TM_2})_{22}]= \frac{1}{6\sqrt{3}}\sin 2\theta \sin\phi, \\
&&I_{1}=\text{Im}[(U_{TM_2})_{11}^*(U_{TM_2})_{12}e^{2 i\alpha}]=\frac{\sqrt{2}}{3}\cos\theta\sin 2\alpha, \\ \nonumber
&&I_{2}=\text{Im}[(U_{TM_2})_{11}^*(U_{TM_2})_{13}e^{2 i\beta}]=\frac{1}{3}\sin 2\theta \sin 2\beta. \\ \nonumber
\nonumber
\end{eqnarray}
The effective Majorana mass $m_{ee}=\left|(M_\nu)_{11}\right|=\left|\sum_{i=1}^{3} (U_{TM_2})_{1i}^2m_i\right|$ is an important physical observable in $0\nu\beta\beta$ decay experiments establishing Majorana nature of neutrinos. In this model, $m_{ee}$ is given by

\begin{eqnarray}\label{mee}
m_{ee}=\frac{1}{3}\left|2m_1\cos^2\theta+m_2 e^{2i\alpha}+2m_3\sin^2\theta e^{2i\beta}\right|.\\ \nonumber
\end{eqnarray}
\begin{table}[t]\label{data}
\begin{center}
\begin{tabular}{l|l|l}
\hline \hline 
Parameter & Best fit $\pm$ \( 1 \sigma \) range & \( 3 \sigma \) range  \\
\hline \multicolumn{2}{c} { Normal mass hierarchy (NH) \( \left(m_{1}<m_{2}<m_{3}\right) \)} \\
\hline $\sin ^{2} \theta_{12} / 10^{-1}$ & $3.18 \pm 0.16$ &  $2.71-3.69$ \\
$\theta_{12} /{ }^{\circ}$ & $34.3 \pm 1.0$  & $31.4-37.4$ \\
$\sin ^{2} \theta_{13} / 10^{-2}$ & $2.200_{-0.062}^{+0.069}$  & $2.000-2.405$ \\
$\theta_{13} /{ }^{\circ}$ & $8.53_{-0.12}^{+0.13}$ & $8.13-8.92$ \\
$\sin ^{2} \theta_{23} / 10^{-1}$ & $5.74 \pm 0.14$ & $4.34-6.10$ \\
$\theta_{23} /{ }^{\circ}$ & $49.26 \pm 0.79$ & $41.20-51.33$ \\
\( \Delta m_{21}^{2}\left[10^{-5} \mathrm{eV}^{2}\right] \) & $7.50_{-0.20}^{+0.22}$& \(6.94-8.14 \) \\
\( |\Delta m_{31}^{2}|\left[10^{-3} \mathrm{eV}^{2}\right] \) & $2.55_{-0.03}^{+0.02}$ & \(2.47-2.63 \) \\
\hline \multicolumn{2}{c} {Inverted mass hierarchy (IH) \( \left(m_{3}<m_{1}<m_{2}\right) \)} \\
\hline $\sin ^{2} \theta_{12} / 10^{-1}$ & $3.18 \pm 0.16$ &  $2.71-3.69$ \\
$\theta_{12} /{ }^{\circ}$ & $34.3 \pm 1.0$  & $31.4-37.4$ \\
$\sin ^{2} \theta_{13} / 10^{-2}$ & $2.225_{-0.070}^{+0.064}$  & $2.018-2.424$ \\
$\theta_{13} /{ }^{\circ}$ & $8.58_{-0.14}^{+0.12}$  & $8.17-8.96$ \\
$\sin ^{2} \theta_{23} / 10^{-1}$ & $5.78_{-0.17}^{+0.10}$ &  $4.33-6.08$ \\
$\theta_{23} /^{\circ}$ & $49.46_{-0.97}^{+0.60}$ & $41.16-51.25$ \\
\( \Delta m_{21}^{2}\left[10^{-5} \mathrm{eV}^{2}\right] \) & $7.50_{-0.20}^{+0.22}$& \(6.94-8.14 \) \\
\( |\Delta m_{31}^{2}|\left[10^{-3} \mathrm{eV}^{2}\right] \) & $2.45_{-0.03}^{+0.02}$ & \( 2.37-2.53\)  \\
\hline 
\end{tabular}
\end{center}
\caption{\label{data}The neutrino oscillation data used in the numerical analysis \cite{deSalas:2020pgw}.}
\end{table}	
The current and forthcoming $0\nu\beta\beta$ decay experiments such as SuperNEMO\cite{Barabash:2011row}, KamLAND-Zen\cite{KamLAND-Zen:2016pfg}, NEXT\cite{NEXT:2009vsd,NEXT:2013wsz} and nEXO\cite{Licciardi:2017oqg} have impressive sensitivity $\mathcal{O}(10^{-2})$ eV to probe this, yet elusive, decay process. Furthermore, the cosmological bound on sum of neutrino masses ($\sum m_{i}$, $i=1,2,3$) is another physical observable which can have imperative implication for the viability of the model. With an increasing statistics, the Planck data Pl18[TT, TE, EE+lowE+lensing] combined with BAO put a very strong upper bound of $0.12$ eV at $95\%$ CL \cite{Planck:2018vyg}. However, we consider relatively stable and conservative bound of 1 eV in the numerical analysis.  

\begin{figure}[t]
\begin{center}
{\epsfig{file=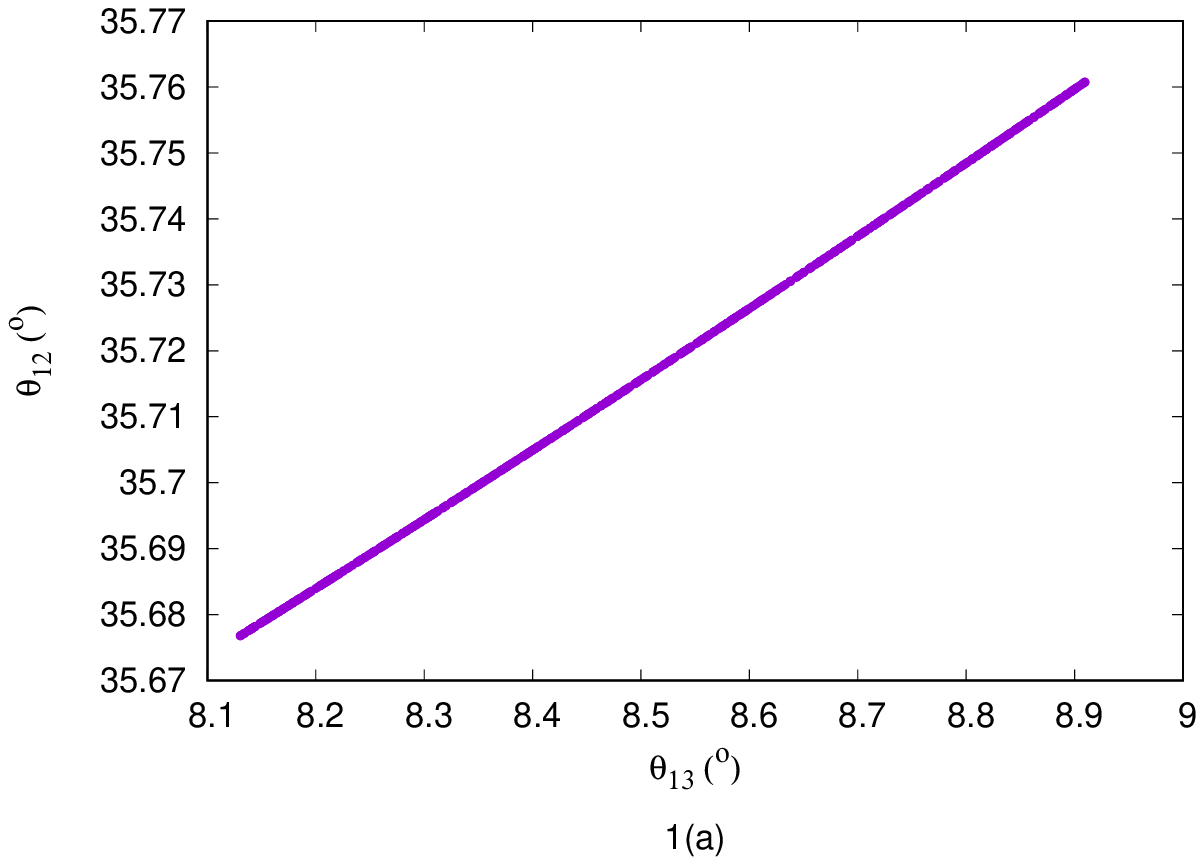,height=6cm,width=6cm}\epsfig{file=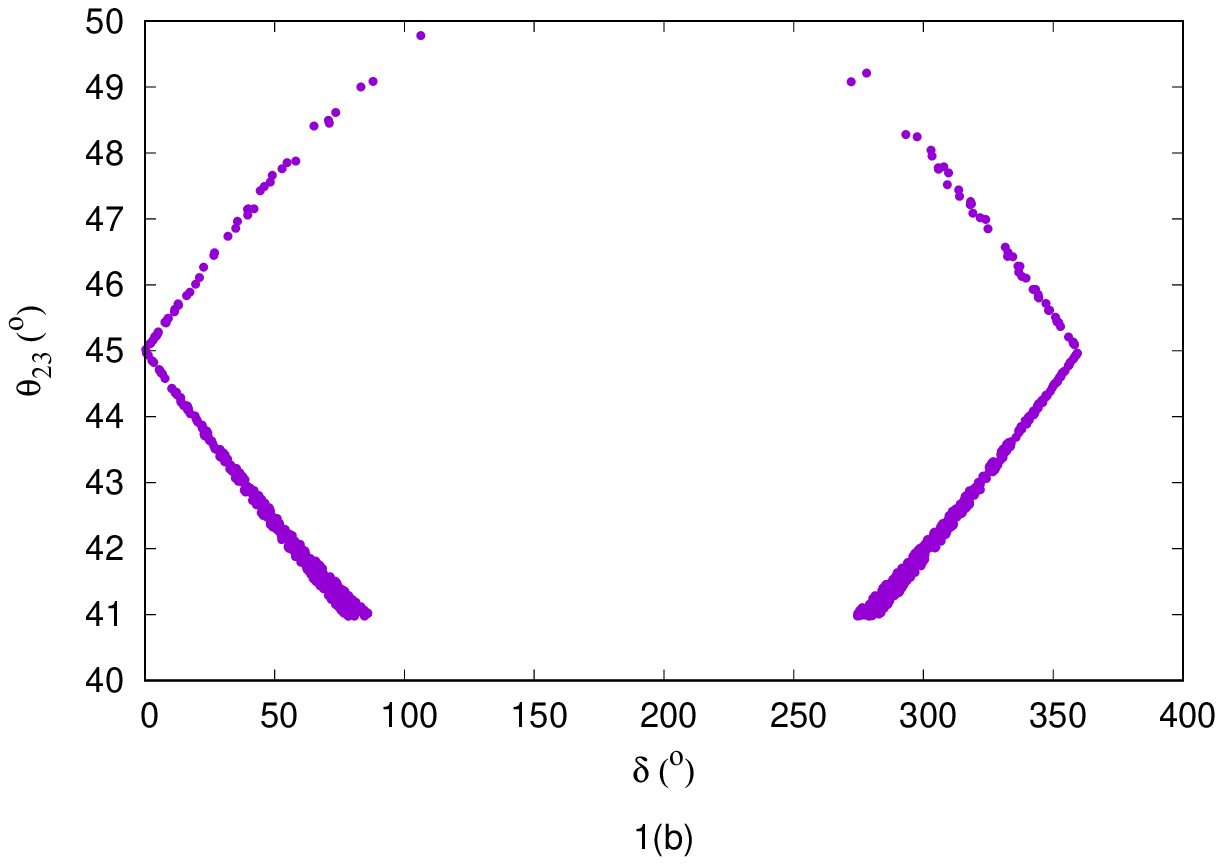,height=6cm,width=6cm}}
{\epsfig{file=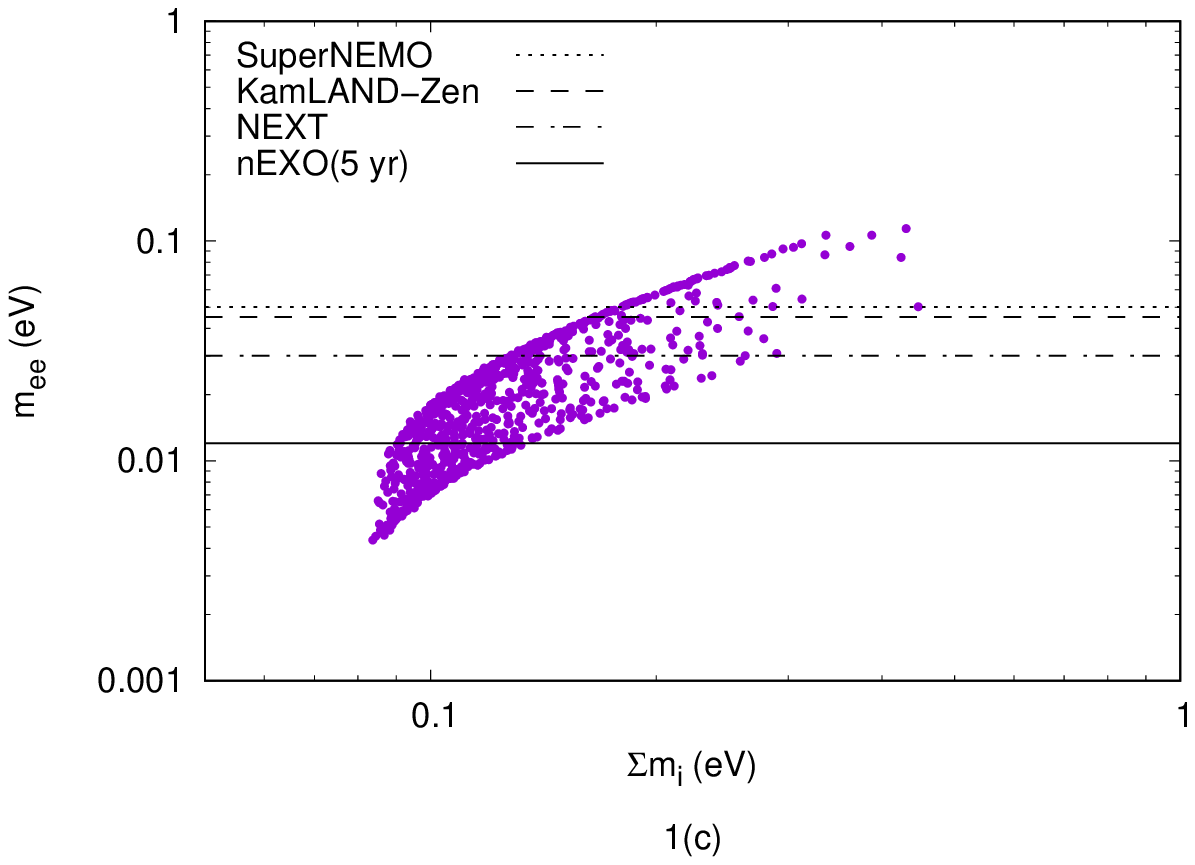,height=6cm,width=6cm}\epsfig{file=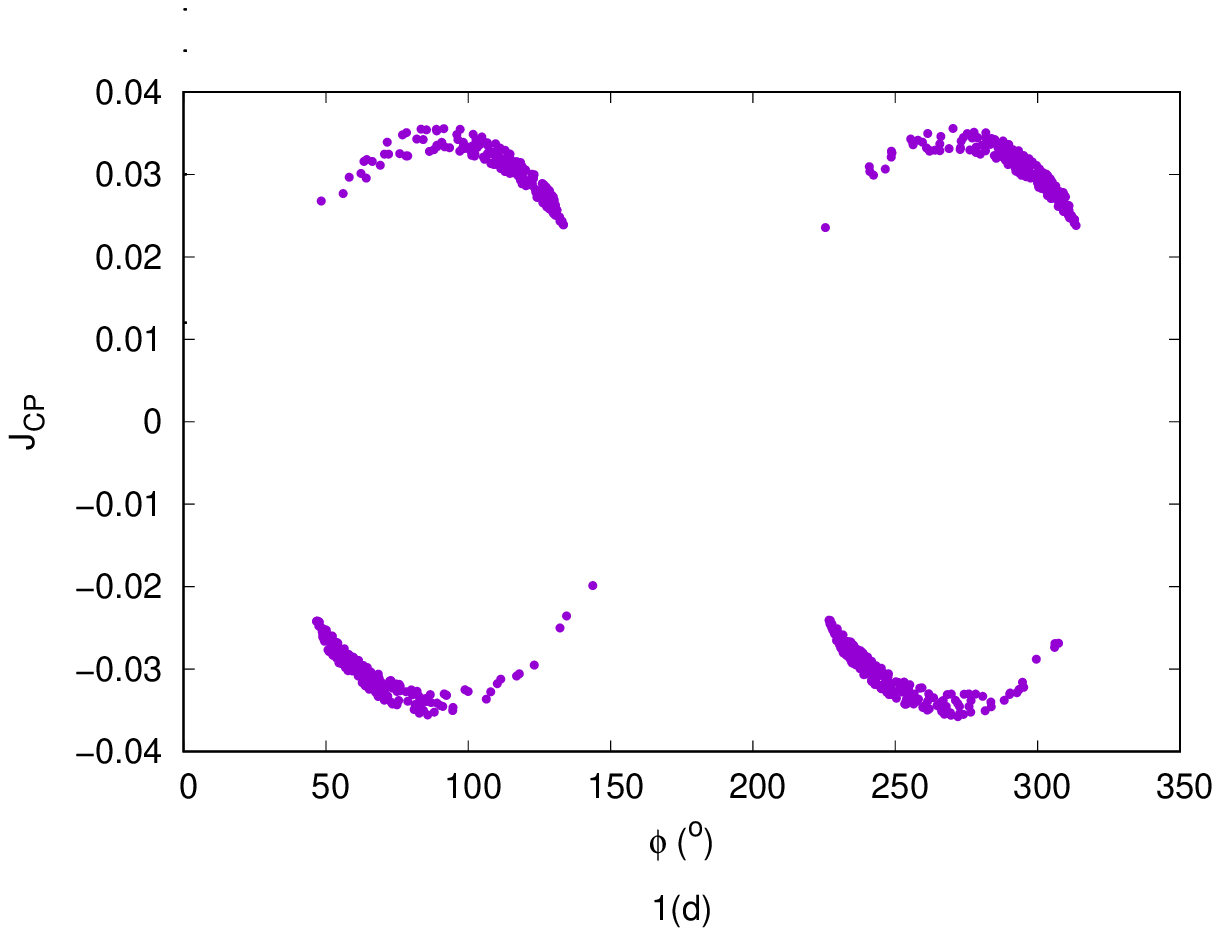,height=6cm,width=6cm}}
{\epsfig{file=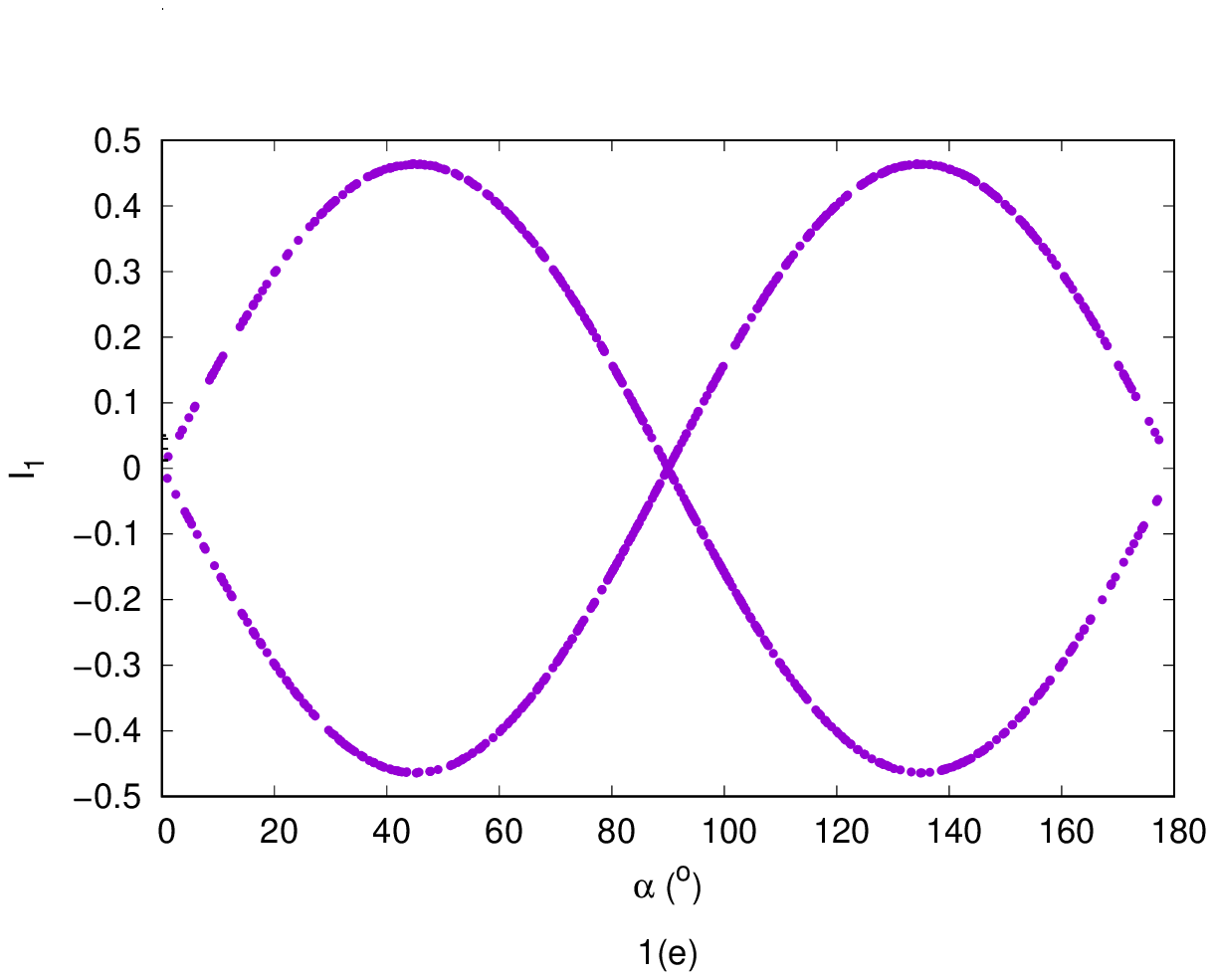,height=6cm,width=6cm}\epsfig{file=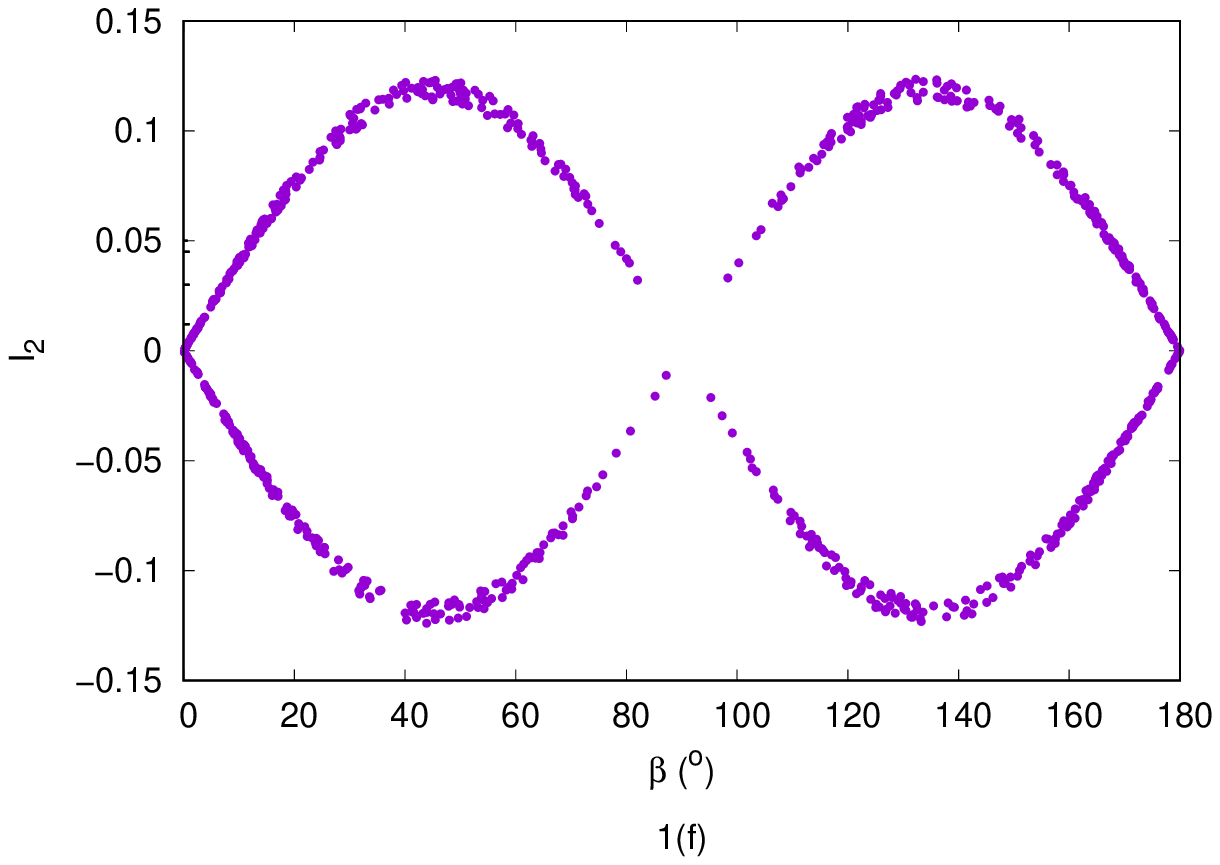,height=6cm,width=6cm}}

\end{center}
\caption{The predictions for TM$_2$ mixing with normal hierarchy of neutrino masses. The horizontal lines in Fig. \ref{fig1}(c) are sensitivities of the respective $0\nu\beta\beta$ decay experiments .}\label{fig1}
\end{figure}

\section{Numerical Analysis and Discussion}
\noindent In the numerical analysis, we have used neutrino oscillation data given in Table \ref{data} and 3$\sigma$ experimental range of $R_\nu$ ($0.025\le R_{\nu}\le0.036$) to constrain the allowed parameter space. In order to obtain mass ratios (Eqns. (\ref{r21}) and (\ref{r31})), the values of free parameters $\theta, \phi, \alpha$ and $\beta$ have been generated randomly with uniform distribution within their physical ranges. Out of $10^9$ 
\begin{figure}[t]
\begin{center}
{\epsfig{file=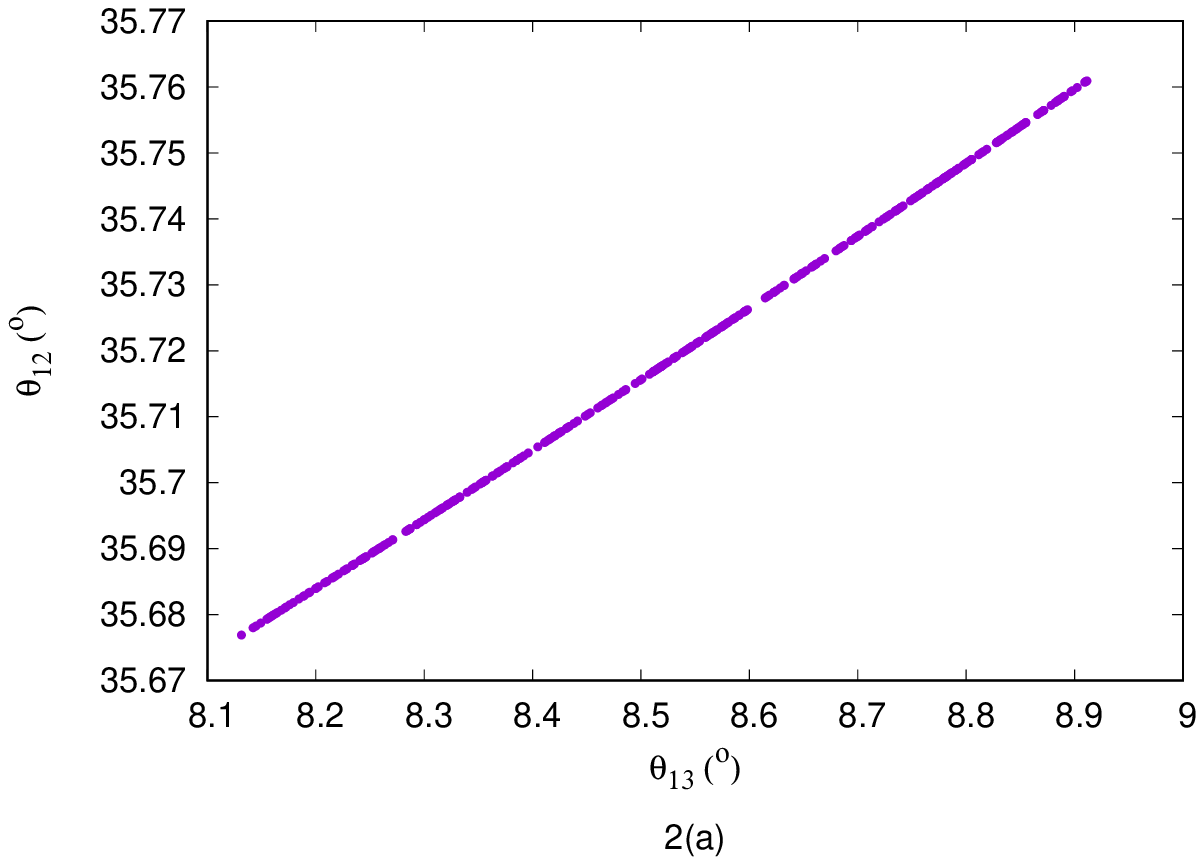,height=6cm,width=6cm}\epsfig{file=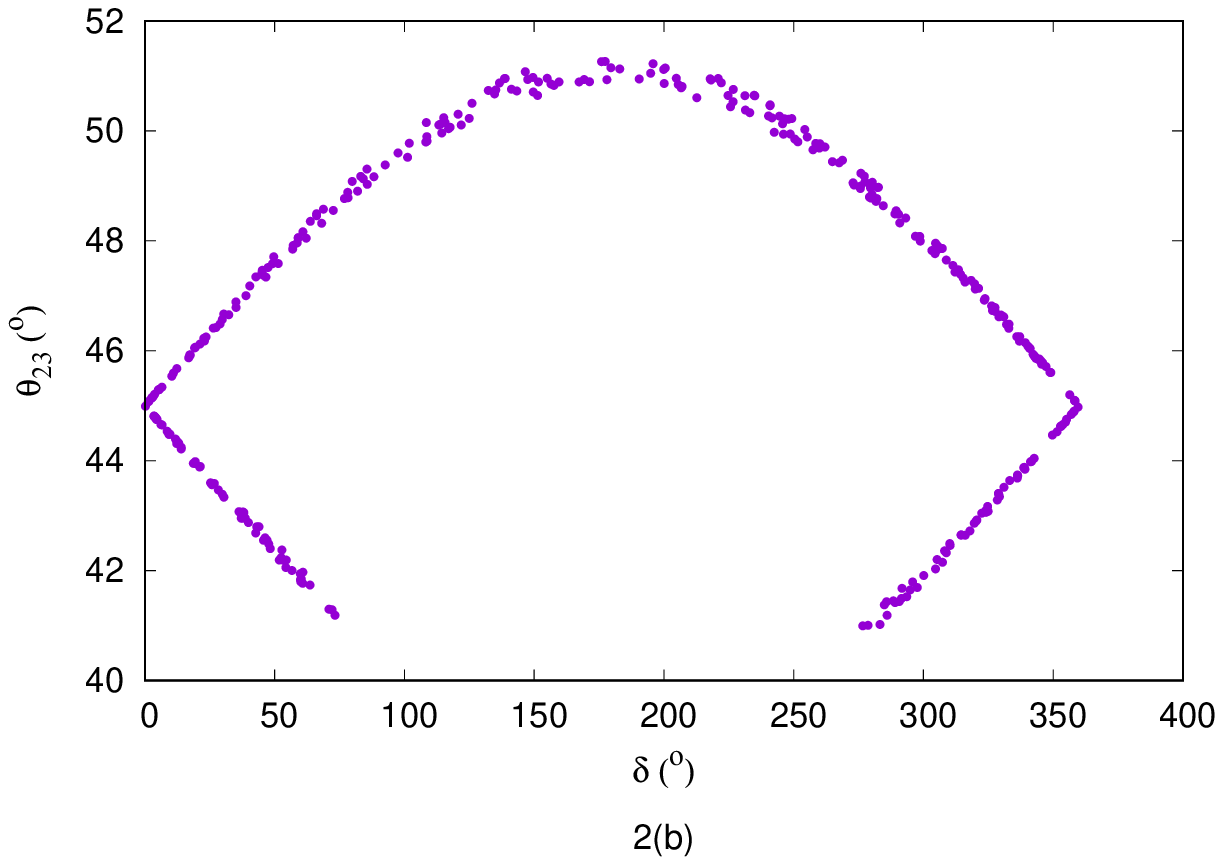,height=6cm,width=6cm}}
{\epsfig{file=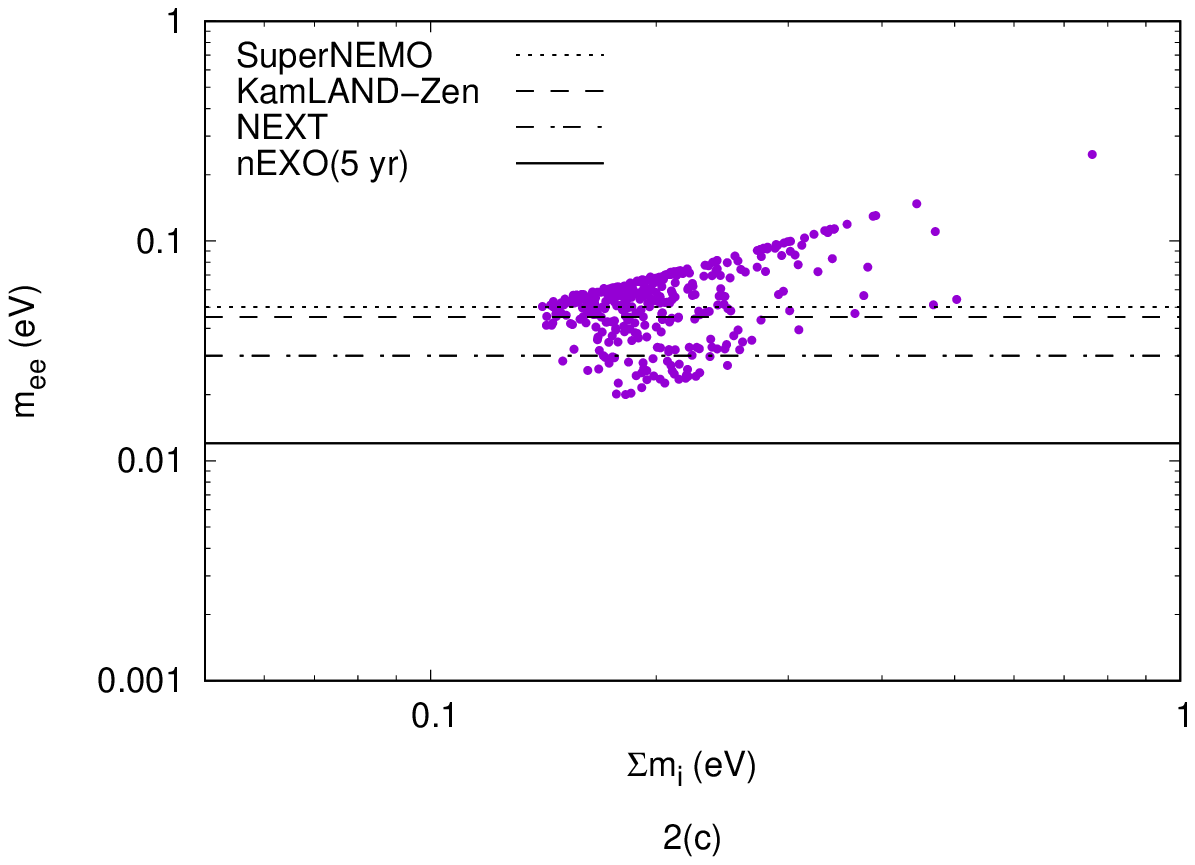,height=6cm,width=6cm}\epsfig{file=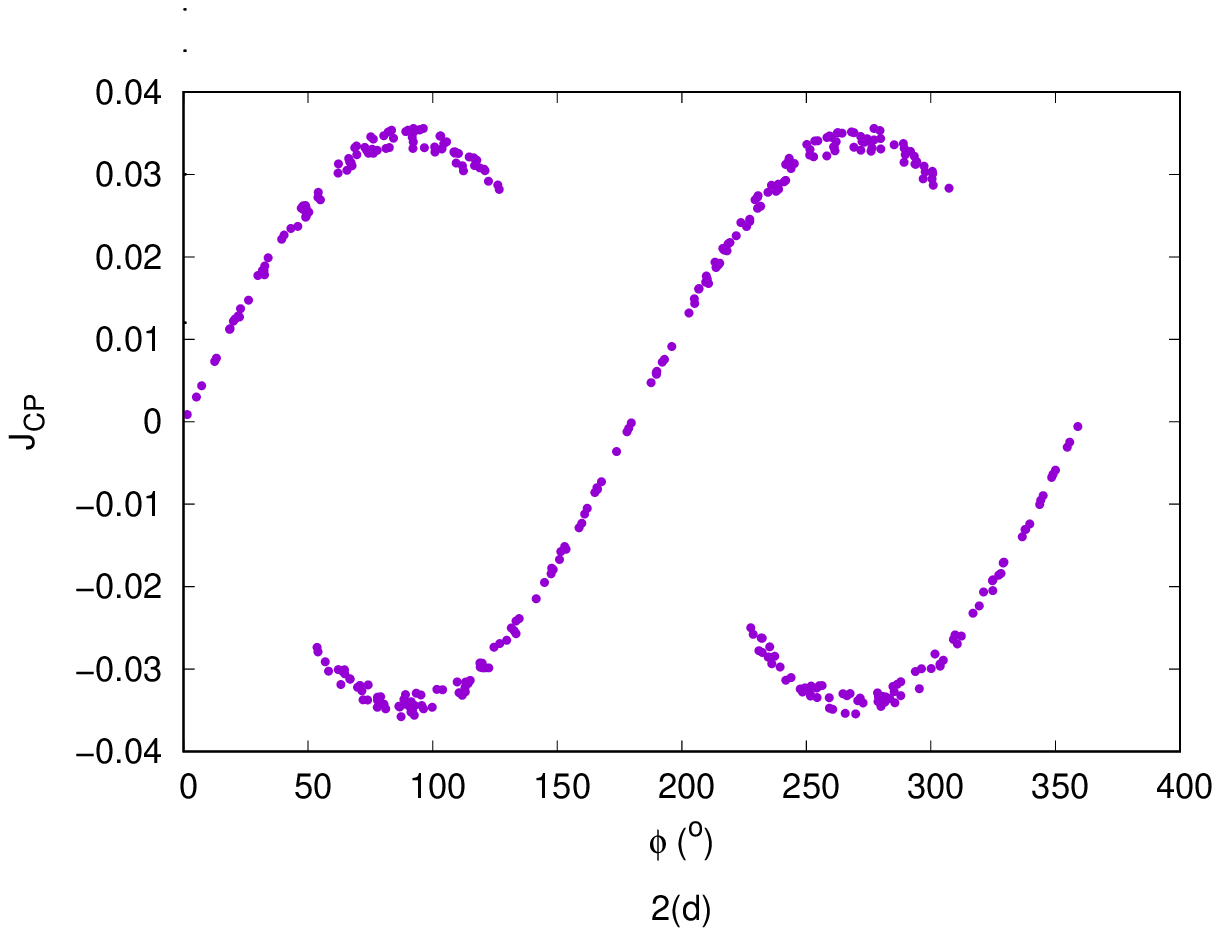,height=6cm,width=6cm}}
{\epsfig{file=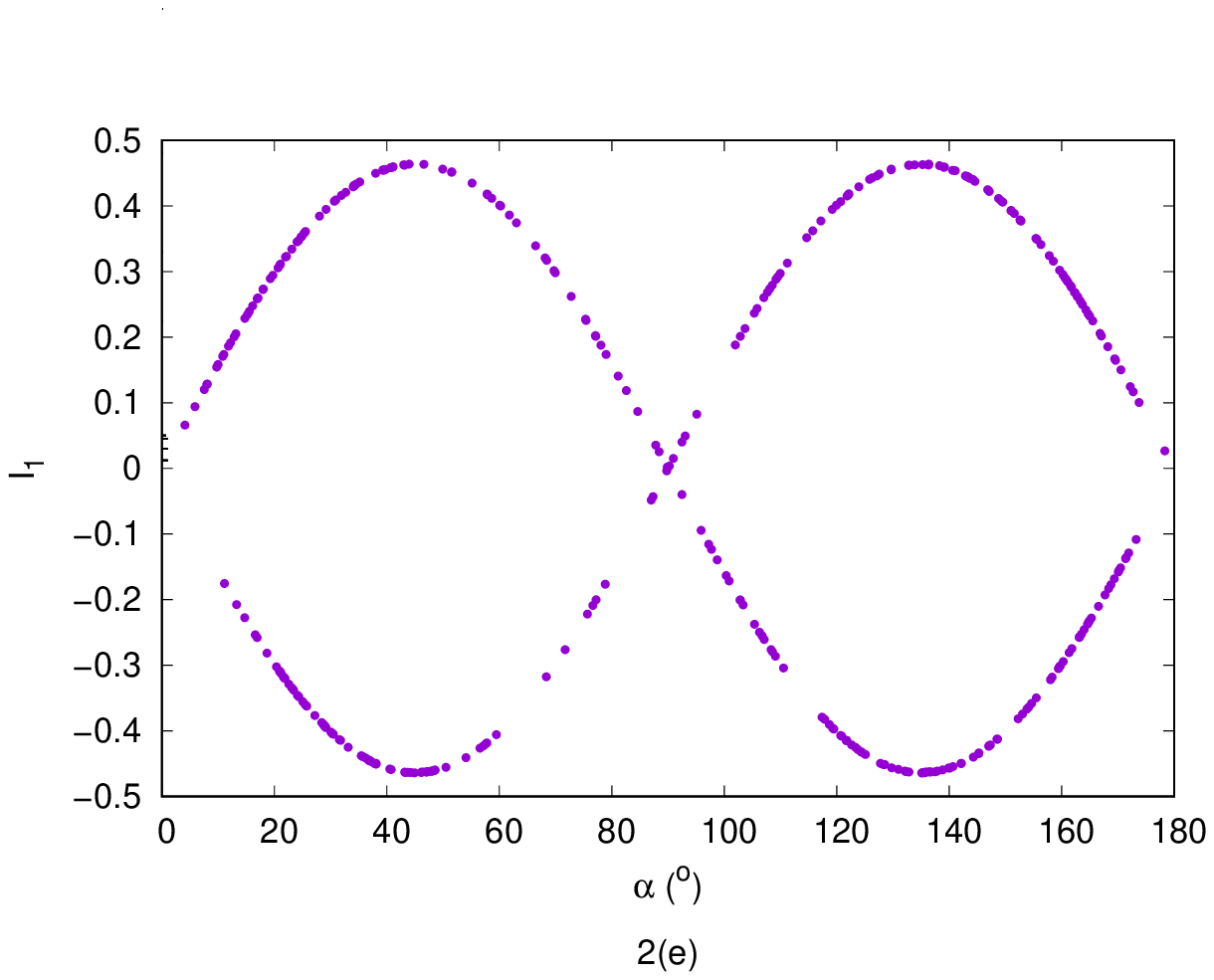,height=6cm,width=6cm}\epsfig{file=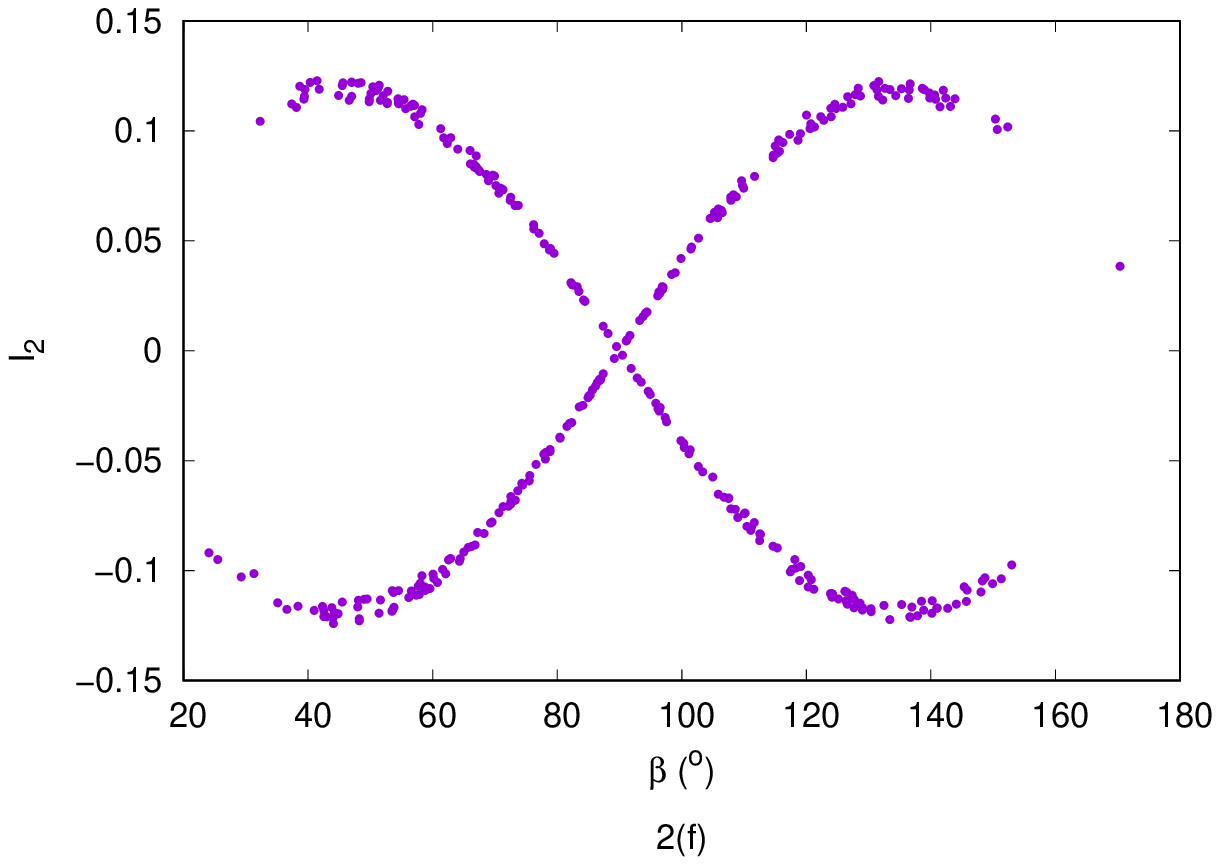,height=6cm,width=6cm}}

\end{center}
\caption{The predictions for TM$_2$ mixing with inverted hierarchy of neutrino masses. The horizontal lines in Fig. \ref{fig2}(c) are sensitivities of the respective $0\nu\beta\beta$ decay experiments.}\label{fig2}
\end{figure}
random samples of the possible solutions only those parameters sets are allowed for which $R_\nu$ (Eqn. (\ref{rnu})) lies within its 3$\sigma$ experimental range. 

\noindent The predictions are depicted as correlation plots amongst different observables in Figs. \ref{fig1} and \ref{fig2}, at 3$\sigma$. In Fig. \ref{fig1} we have shown predictions for TM$_2$ mixing with normal hierarchy (NH) of neutrino masses. One of the generic feature of TM$_2$ mixing is that there exist a slight tension in $\theta_{12}$ prediction at 1$\sigma$ which is, also, exhibited in Fig. \ref{fig1}(a) (experimental range of $\theta_{12}$, at 1$\sigma$, is $33.3^o<\theta_{12}<35.3^o$). $\theta_{23}$, in general, can be in upper ($\theta_{23}>45^o$) or lower ($\theta_{23}<45^o$) octant including its maximal value ($\theta_{23}=45^o$). Furthermore, it is evident from Fig. \ref{fig1}(b) that $CP$ violating phase $\delta$ lies in first and fourth quadrants. The $CP$ conserving solutions (with $\delta=0^o$ or $360^o$) require a maximal value of  $\theta_{23}$. Thus, in order to have $CP$ violation measurable in neutrino oscillation experiments $\theta_{23}$ must deviate from allowed maximal value. It is evident from Fig. \ref{fig1}(c) that there exist allowed region of effective Majorana mass parameter $m_{ee}$ which is beyond the sensitivity reach of $0\nu\beta\beta$ decay experiments. The predicted lower bound on $m_{ee}$ is 0.004 eV, at 3$\sigma$. The model is, in general, consistent with both $CP$ conserving and $CP$ violating solutions (Figs. \ref{fig1}(d-f)). The allowed ranges of the parameters, at 3$\sigma$, are given in Table \ref{tab2}.

\noindent In Fig. \ref{fig2}, we depict the correlation plots of the parameters assuming inverted hierarchical (IH) spectrum of neutrino masses. As in Fig. \ref{fig1}(a), Fig. \ref{fig2}(a), also, shows a positive linear correlation between $\theta_{13}$ and $\theta_{12}$ with mild tension in predicted values of $\theta_{12}$ at 1$\sigma$. For $\theta_{23}$ below maximality ($\theta_{23}<45^o$), Dirac $CP$ phase $\delta$ can be in first ($0^o<\delta<90^o$) and fourth ($270^o<\delta<360^o$) quadrants, however, above maximality ($\theta_{23}>45^o$), $\delta$ remains unconstrained i.e. allowed in whole physical range ($0^o<\delta<360^o$). Also, in order to have $CP$ violation in the leptonic sector $\theta_{23}$ must not have maximal value i.e.  $\theta_{23}\ne 45^o$. The effective Majorana mass $m_{ee}$ is bounded from below $m_{ee}>0.02$ eV which can be probed in $0\nu\beta\beta$ decay experiments. The non-observation of $0\nu\beta\beta$ decay or invoking the cosmological bound of $0.12$ eV on $\sum m_i$, at 95\% CL, shall refutes inverted hierarchy implying TM$_2$ mixing with normal hierarchy as the only viable possibility in the model. The $CP$ invariants $J_{CP}$, $I_1$ and $I_2$ are found to be in the ranges ($-0.035-0.035$), ($-0.460-0.460$) and ($-0.130-0.130$) at 3$\sigma$, respectively (Fig. \ref{fig2}(d-f)). In the following, we have constructed a flavor model which realizes the extended magic symmetry ansatz considered in this work.

\begin{table}[t]
\centering
\begin{tabular}{|l|l|l|}
\hline\hline
Parameter       & TM$_2$ mixing with NH                    & TM$_2$ mixing with IH                                                                                                                                                                                                       \\ \hline
$m_{ee}$ (eV)  & \textgreater{}0.004                      & \textgreater{}0.02                                                                                                                                                                                                           \\ \hline
$J_{CP}$        & \begin{tabular}[c]{@{}l@{}}(0.024$-$0.035)$\oplus$\\ (-0.035$-$-0.018)\end{tabular} & (-0.035$-$0.035)                                                                                                                                                                                        \\ \hline
I$_1$           & (-0.470$-$0.470)                         & (-0.460$-$0.460)                                                                                                                                                                                                               \\ \hline
I$_2$           & (-0.130$-$0.130)                         & (-0.130$-$0.130)                                                                                                                                                                                                                \\ \hline
$\delta$ ($^o$) & \begin{tabular}[c]{@{}l@{}}(0$-$90)$\oplus$\\ (270$-$360)\end{tabular}             & \begin{tabular}[c]{@{}l@{}}(0$-$90)$\oplus$(270$-$360)\\ for $\theta_{23}<45^o$\end{tabular}  \\ \hline
$\sum m_i$ (eV) & (0.08$-$0.35) & \begin{tabular}[c]{@{}l@{}}(0.15$-$0.80),\\ in light of \\ cosmological bound\\ $\sum m_i<0.12$ eV, \\ this scenario is\\ disallowed\end{tabular} \\ \hline
\end{tabular}
\caption{\label{tab2}  The numerical predictions of the model at 3$\sigma$.}
\end{table}

\section{Symmetry Realization of Extended Magic Symmetry based on $\Delta (54)$ Group}
\noindent In the effective field theory framework, we obtain the Type-I+II seesaw origin of magic symmetry. The standard model (SM)  fermionic field content is enlarged by a Majorana fermion right-handed neutrino ($N_{1}$) while scalar field content comprises of scalar triplet ($\Delta$) and flavon fields ($\rho_{1},\rho_{2},\chi,\Phi_{i}(i=1,2,3)$). $\Delta(54)$ symmetry have been employed to obtain the Yukawa terms relevant for neutrino mass generation through the magic neutrino mass matrix. $\Delta(54)$ is a non-Abelian discrete symmetry having ten irreducible representations $viz.$, two singlets ($1_{1},1_{2}$), four doublets ($2_{1},2_{2},2_{3},2_{4}$) and four triplets ($3_{1(1)},3_{1(2)},3_{2(1)},3_{2(2)}$) \cite{Ishimori:2010au, Loualidi:2021qoq}. The tensor products of irreducible representations can be decomposed as shown in Appendix \ref{appendix}.

\begin{table}[t]
		\centering
		\begin{tabular}{cccccccccccc}
			Symmetry & $D_{iL}$  & $l_{R}$ & $H$ &  $\Delta$ & $N_{1}$ & $\rho_{1}$ &$\rho_{2}$& $\chi$ & $\Phi_{1}$&$\Phi_{2}$ &$\Phi_{3}$ \\ \hline
			$SU(2)_L$    &     2 &     1 &     2      &    3      &   1     &   1& 1       & 1 & 1 &    1     & 1  \\ \hline
			$\Delta(54)$ &      $3_{2(1)}$ &$3_{1(2)}$      &     $1_{1}$    &      $1_{2}$    &    $1_{1}$      &  $1_{2}$ & $1_{2}$ & $2_{1}$ & $3_{1(1)}$ & $3_{2(2)}$ & $3_{2(1)}$    \\ \hline
			$Z_{4}$ & 1 & $i$ & 1 & $-i$ & -1 & 1 & $-i$& $-i$&$i$& -1 &  $i$ \\ \hline
			$Z_{5}$ & 1 & $\eta^2$ & 1 & $\eta^{4}$ & 1 & 1 & $\eta^3$& $\eta^3$&$\eta$& 1 &  $\eta$ \\
			
		\end{tabular}
		\caption{\label{tab3} The field content of the model and respective charge assignments under $SU(2)_L$, $\Delta(54)$, $Z_4$ and $Z_5$ symmetries, where $\eta=e^{\frac{2\pi i}{5}}$.} 
	\end{table}
	
\noindent The charge assignments under $\Delta(54)$, $Z_4$ and $Z_5$ are given in Table \ref{tab3}. Here, $Z_4$ and $Z_5$ Abelian symmetries have been employed to constrain Yukawa structure of the Lagrangian. Using the tensor decomposition of $\Delta(54)$, the Lagrangian relevant for charged lepton masses is given by  
\begin{eqnarray}\label{lcharged}
\nonumber
\mathcal{L}_{C}&&= \alpha(\bar{D}_{iL} H l_{R} ) \rho_{2}+\beta(\bar{D}_{i L} H l_{R}) \chi+H.c.,\\
\nonumber
&&=\alpha(\bar{D}_{eL} e_{R}+\bar{D}_{\mu L} \mu_{R}+ \bar{D}_{\tau L} \tau_{R} ) H \rho_{2}\\
&& +\beta\left[(\omega \bar{D}_{eL} e_{R}+\omega^{2} \bar{D}_{\mu L} \mu_{R}+ \bar{D}_{\tau L} \tau_{R})\chi_{1}-(\bar{D}_{eL} e_{R}+\omega^{2} \bar{D}_{\mu L} \mu_{R}+ \omega \bar{D}_{\tau L} \tau_{R})\chi_2\right]+H.c.,
\end{eqnarray} 
where $\alpha$ and $\beta$ are coupling constants. After spontaneous symmetry breaking, the flavon fields acquire $vevs$, $\langle \rho_{2}\rangle_{0}=v_{2}$ and $\langle \chi(\chi_{1},\chi_{2})\rangle_{0}=(v_{\chi_1},v_{\chi_2})$. The first term in Eqn. (\ref{lcharged}) results in diagonal charged lepton mass matrix with degenerate masses while the second term breaks the degeneracy leading to non-degenerate charged lepton masses. The charged lepton mass matrix is given by 
\begin{equation} \label{mcharged}
	M_{l}=
	\frac{v_{H}v_{2}}{\sqrt{2}}{\begin{pmatrix}
	\alpha & 0 & 0\\
  0& \alpha  &0 \\
    0 & 0 & \alpha 
		\end{pmatrix}}+
	\frac{\beta v_{H}}{\sqrt{2}}{\begin{pmatrix}
\omega v_{\chi_1}-v_{\chi_2} & 0 & 0\\
  0& \omega^{2} v_{\chi_1}-\omega^{2} v_{\chi_2}  &0 \\
    0 & 0 & v_{\chi_1} -\omega  v_{\chi_2}  
		\end{pmatrix}},
	\end{equation}
where $v_{H}/\sqrt{2}$ is $vev$ of the neutral component of Higgs field ($H$). \\
\underline{\textbf{Type-I seesaw:}} At dimension-5, using the flavon field ($\Phi_2$) and one right-handed neutrino ($N_1$), Type-I seesaw mechanism is implemented via Dirac mass matrix ($M_{D}$) and right-handed neutrino mass matrix($M_{R}$). The relevant Lagrangian is given by
\begin{equation}\label{type1}
    \mathcal{L}_{I}=y_{D}(D_{iL}\Tilde{H}N_{1})\Phi_{2}+ M \bar{N}_{1}^{c}N_{1}+H.c.,
    \end{equation}
    where $\Tilde{H}=i\tau_{2}H$, $\tau_{2}$ is Pauli matrix and $M$ is bare mass term for $N_1$. The mass matrices $M_{D}$ and $M_{R}$ are given by 
    \begin{equation} \label{dirac}
	M_{D}=
	{\begin{pmatrix}
	y_{D} \phi_{2}^{(1)} \\
     y_{D} \phi_{2}^{(2)} \\
    y_{D} \phi_{2}^{(3)}
		\end{pmatrix}},\hspace{0.2cm}
		M_{R}=M.
	\end{equation}
With vacuum alignment $v_{\phi_{2}}$($0,-1,1$) of the Flavon field $\Phi_{2}$,   Type-I seesaw contribution to $M_\nu$ is
\begin{eqnarray}\label{mnu1}
\nonumber
M_{\nu}^I&&=-M_{D}M_{R}^{-1}M_{D}^{T}\\
&&=\begin{pmatrix}
	0 &0&0\\
    0& b &-b \\
    0 &-b &b
		\end{pmatrix},
\end{eqnarray}
where  $b=y_{D}^{2}v_{\phi_2}^{2}/M$.

\noindent \underline{\textbf{Type-II seesaw:}} Furthermore, the addition of scalar triplet field ($\Delta$) leads to Type-II seesaw contribution to $M_\nu$. The Yukawa Lagrangian responsible for this contribution is given by
\begin{eqnarray}\label{typeii}
\nonumber
\mathcal{L}_{II}&&= y_{\Delta_1}(\overline{D^{c}_{iL}}i\tau_{2} \Delta D_{jL}) \Phi_{1}+y_{\Delta_2}(\overline{D^{c}_{iL}}i\tau_{2}\Delta D_{jL}) \rho_{1} \Phi_{1}+y_{\Delta_3}(\overline{D^{c}_{iL}}i\tau_{2} \Delta D_{jL}) \Phi_{3}+H.c.,\\
\nonumber
&&=y_{\Delta_1}(\overline{D^{c}_{eL}}  D_{eL} +\overline{D^{c}_{\mu L}}  D_{\mu L}+\overline{D^{c}_{\tau L}}  D_{\tau L})v_{\Delta}v_{\phi_{1}} \\
\nonumber 
&&+y_{\Delta_2}(\overline{D^{c}_{eL}} D_{\tau L} +\overline{D^{c}_{\mu L}}  D_{\mu L}+\overline{D^{c}_{\tau L}}  D_{e L})v_{\rho}v_{\Delta}v_{\phi_{1}}\\
&& +y_{\Delta_3}(\overline{D^{c}_{eL}}  D_{\mu L} +\overline{D^{c}_{\mu L}}  D_{e L}+\overline{D^{c}_{\tau L}}  D_{\tau L})v_{\Delta}v_{\phi_{3}}+H.c.,
\end{eqnarray} 
with the flavon alignments $v_{\phi_{1}}$($0,1,0$) and $v_{\phi_{3}}$($0,0,1$) for the fields $\Phi_1$ and $\Phi_3$, respectively; $v_{\Delta}$ is the $vev$ of scalar triplet field, $\Delta$.
The Type-II seesaw contribution to neutrino mass matrix is given by
\begin{equation}\label{type2}
M_{\nu}^{II}=\begin{pmatrix}
	a & d & c\\
    d & a+c & 0 \\
    c & 0 & a+d
		\end{pmatrix},
\end{equation}
where $a=y_{\Delta_1}v_{\phi_{1}}v_{\Delta}$, $c=y_{\Delta_2}v_{\phi_{1}}v_{\Delta}v_{\rho_1}$ and $d=y_{\Delta_3}v_{\phi_{3}}v_{\Delta}$.
\noindent Within Type-I+II seesaw setting, with $d=b$, the total effective Majorana neutrino mass matrix is given by
\begin{eqnarray}\label{totalmnu}
\nonumber
    M_{\nu}&&=M_{\nu}^{I}+M_{\nu}^{II},\\
    &&=\begin{pmatrix}
	a & b & c\\
    b & a+b+c & -b \\
    c & -b & 2b+a
		\end{pmatrix},
\end{eqnarray}
which is the magic neutrino mass matrix with (2,2) element equals to \textit{``magic sum"} i.e. $a+b+c$ (Eqn. (\ref{exmnu})). This is a representative realization of the phenomenological ansatz considered in this work. For completeness of the model, the flavon alignments of fields $\Phi_{i}(i=1,2,3)$ are required to be tested under the minimization of scalar potential\cite{Altarelli:2005yp,Altarelli:2010gt,King:2013eh}.

\section{Conclusions}
\noindent In conclusion, we have considered an extension of the magic symmetry ansatz within the paradigm of TM$_2$ mixing scheme wherein (2,2) element of $M_\nu$ is, also, equal to the \textit{``magic sum"}. This phenomenological ansatz exhibits strong correlations and homoscedasticity  amongst the physical observables signifying the high predictability of the model. Some interesting phenomenological consequences of the model are:

\begin{itemize}
    \item In order to have $CP$ violation measurable in the neutrino oscillation experiments, the atmospheric mixing angle $\theta_{23}$ is predicted to be non-maximal i.e. for $\theta_{23}\ne45^o$.
  
    \item The observables $m_{ee}$ and $\sum m_i$ have imperative implication for the model. There exist a lower bound on $m_{ee}>0.02$ eV for TM$_2$ (with IH) which can be probed in $0\nu\beta\beta$ decay experiments. The non-observation of $0\nu\beta\beta$ decay shall refute IH for TM$_2$ mixing. 
    \item Though we have considered a relatively stable and conservative bound, on $\sum m_i$,  of 1 eV in the numerical analysis but the current bound from Planck data combined with WMAP and BAO of 0.12 eV at 95\% CL refutes IH in the model (Fig. \ref{fig2}(c)) implying TM$_2$ with normal hierarchy as
the only viable possibility in the model. The numerical predictions of the model, at 3$\sigma$, are given in Table \ref{tab2}.
    \item The model parameter space is, further, constrained if we take into consideration the recent global reassessment of the neutrino oscillation parameters\cite{deSalas:2020pgw} which have reported Dirac type $CP$ phase $\delta$ in the ranges ($128^o-359^o$) and ($200^o-353^o$), at 3$\sigma$, for normal and inverted hierarchical neutrino masses, respectively. For example, region of the parameter space for which $\delta$ lies in the first quadrant is disallowed (see Figs. \ref{fig1}(b) and \ref{fig2}(b)). 
\end{itemize}     

\noindent We have, also, proposed a scenario, based on $\Delta (54)$ symmetry within Type-I+II seesaw setting to realize the extended magic symmetry ansatz, considered in this work.

\noindent\textbf{\Large{Acknowledgments}}
 \vspace{.3cm}\\
 M. K. acknowledges the financial support provided by Department of Science and Technology (DST), Government of India vide Grant No. DST/INSPIRE Fellowship/2018/IF180327. The authors, also, acknowledge Department of Physics and Astronomical Science for providing necessary facility to carry out this work.

\section*{Declarations} The authors declare that they have no known competing financial interests or personal relationships that could have appeared to influence the work reported in this paper.

\begin{appendices}
\section{Tensor products of $\Delta(54)$}\label{appendix}
$$
\mathbf{1_{1}}\otimes \mathbf{S_{i}}=\mathbf{S_{i}}, \hspace{0.5cm} \mathbf{1_{2}}\otimes\mathbf{1_{2}}=\mathbf{1_{1}}, \hspace{0.5cm} \mathbf{1_{2}}\otimes\mathbf{3_{1(1)}}=\mathbf{3_{2(1)}},
$$
$$
\mathbf{1_{2}}\otimes\mathbf{3_{1(2)}}=\mathbf{3_{2(2)}}, \hspace{0.5cm} \mathbf{1_{2}}\otimes\mathbf{3_{2(1)}}=\mathbf{3_{1(1)}}, \hspace{0.5cm} \mathbf{1_{2}}\otimes\mathbf{3_{2(2)}}=\mathbf{3_{1(2)}}
$$

$$
\left(\begin{array}{l}
a_{1} \\
a_{2} \\
a_{3}
\end{array}\right)_{\mathbf{3_{1(1)}}} \otimes\left(\begin{array}{l}
b_{1} \\
b_{2} \\
b_{3}
\end{array}\right)_{\mathbf{3_{1(1)}}}=\left(\begin{array}{l}
a_{1} b_{1} \\
a_{2} b_{2} \\
a_{3} b_{3}
\end{array}\right)_{\mathbf{3_{1(2)}}} \oplus\left(\begin{array}{l}
a_{2} b_{3}+a_{3} b_{2} \\
a_{3} b_{1}+a_{1} b_{3} \\
a_{1} b_{2}+a_{2} b_{1}
\end{array}\right)_{\mathbf{3_{1(2)}}} \oplus\left(\begin{array}{l}
a_{2} b_{3}-a_{3} b_{2} \\
a_{3} b_{1}-a_{1} b_{3} \\
a_{1} b_{2}-a_{2} b_{1}
\end{array}\right)_{\mathbf{3_{2(2)}}},
$$

$$
\left(\begin{array}{l}
a_{1} \\
a_{2} \\
a_{3}
\end{array}\right)_{\mathbf{3_{1(2)}}} \otimes\left(\begin{array}{l}
b_{1} \\
b_{2} \\
b_{3}
\end{array}\right)_{\mathbf{3_{1(2)}}}=\left(\begin{array}{l}
a_{1} b_{1} \\
a_{2} b_{2} \\
a_{3} b_{3}
\end{array}\right)_{\mathbf{3_{1(1)}}} \oplus\left(\begin{array}{l}
a_{2} b_{3}+a_{3} b_{2} \\
a_{3} b_{1}+a_{1} b_{3} \\
a_{1} b_{2}+a_{2} b_{1}
\end{array}\right)_{\mathbf{3_{1(1)}}} \oplus\left(\begin{array}{l}
a_{2} b_{3}-a_{3} b_{2} \\
a_{3} b_{1}-a_{1} b_{3} \\
a_{1} b_{2}-a_{2} b_{1}
\end{array}\right)_{\mathbf{3_{2(1)}}},
$$

$$
\left(\begin{array}{l}
a_{1} \\
a_{2} \\
a_{3}
\end{array}\right)_{\mathbf{3_{2(1)}}} \otimes\left(\begin{array}{l}
b_{1} \\
b_{2} \\
b_{3}
\end{array}\right)_{\mathbf{3_{2(1)}}}=\left(\begin{array}{l}
a_{1} b_{1} \\
a_{2} b_{2} \\
a_{3} b_{3}
\end{array}\right)_{\mathbf{3_{1(2)}}} \oplus\left(\begin{array}{l}
a_{2} b_{3}+a_{3} b_{2} \\
a_{3} b_{1}+a_{1} b_{3} \\
a_{1} b_{2}+a_{2} b_{1}
\end{array}\right)_{\mathbf{3_{1(2)}}} \oplus\left(\begin{array}{l}
a_{2} b_{3}-a_{3} b_{2} \\
a_{3} b_{1}-a_{1} b_{3} \\
a_{1} b_{2}-a_{2} b_{1}
\end{array}\right)_{\mathbf{3_{2(2)}}},
$$

$$
\left(\begin{array}{l}
a_{1} \\
a_{2} \\
a_{3}
\end{array}\right)_{\mathbf{3_{2(2)}}} \otimes\left(\begin{array}{l}
b_{1} \\
b_{2} \\
b_{3}
\end{array}\right)_{\mathbf{3_{2(2)}}}=\left(\begin{array}{l}
a_{1} b_{1} \\
a_{2} b_{2} \\
a_{3} b_{3}
\end{array}\right)_{\mathbf{3_{1(1)}}} \oplus\left(\begin{array}{l}
a_{2} b_{3}+a_{3} b_{2} \\
a_{3} b_{1}+a_{1} b_{3} \\
a_{1} b_{2}+a_{2} b_{1}
\end{array}\right)_{\mathbf{3_{1(1)}}} \oplus\left(\begin{array}{l}
a_{2} b_{3}-a_{3} b_{2} \\
a_{3} b_{1}-a_{1} b_{3} \\
a_{1} b_{2}-a_{2} b_{1}
\end{array}\right)_{\mathbf{3_{2(1)}}},
$$

$$
\left(\begin{array}{l}
a_{1} \\
a_{2} \\
a_{3}
\end{array}\right)_{\mathbf{3_{1(1)}}} \otimes\left(\begin{array}{l}
b_{1} \\
b_{2} \\
b_{3}
\end{array}\right)_{\mathbf{3_{1(2)}}}=\left(a_{1} b_{1}+a_{2} b_{2}+a_{3} b_{3}\right)_{\mathbf{1_{1}}} \oplus\left(\begin{array}{l}
a_{1} b_{1}+\omega^{2} a_{2} b_{2}+\omega a_{3} b_{3} \\
\omega a_{1} b_{1}+\omega^{2} a_{2} b_{2}+a_{3} b_{3}
\end{array}\right)_{\mathbf{2_{1}}}
$$

$$
\begin{aligned}
&\oplus\left(\begin{array}{l}
a_{1} b_{2}+\omega^{2} a_{2} b_{3}+\omega a_{3} b_{1} \\
\omega a_{1} b_{3}+\omega^{2} a_{2} b_{1}+a_{3} b_{2}
\end{array}\right)_{\mathbf{2_{2}}} \\
&\oplus\left(\begin{array}{l}
a_{1} b_{3}+\omega^{2} a_{2} b_{1}+\omega a_{3} b_{2} \\
\omega a_{1} b_{2}+\omega^{2} a_{2} b_{3}+a_{3} b_{1}
\end{array}\right)_{\mathbf{2_{3}}} \\
&\oplus\left(\begin{array}{l}
a_{1} b_{3}+a_{2} b_{1}+a_{3} b_{2} \\
a_{1} b_{2}+a_{2} b_{3}+a_{3} b_{1}
\end{array}\right)_{\mathbf{2_{4}}},
\end{aligned}
$$

$$
\left(\begin{array}{l}
a_{1} \\
a_{2} \\
a_{3}
\end{array}\right)_{\mathbf{3_{1(1)}}} \otimes\left(\begin{array}{l}
b_{1} \\
b_{2} \\
b_{3}
\end{array}\right)_{\mathbf{3_{2(1)}}}=\left(\begin{array}{l}
a_{1} b_{1} \\
a_{2} b_{2} \\
a_{3} b_{3}
\end{array}\right)_{\mathbf{3_{2(2)}}} \oplus\left(\begin{array}{l}
a_{3} b_{2}-a_{2} b_{3} \\
a_{1} b_{3}-a_{3} b_{1} \\
a_{2} b_{1}-a_{1} b_{2}
\end{array}\right)_{\mathbf{3_{1(2)}}} \oplus\left(\begin{array}{l}
a_{3} b_{2}+a_{2} b_{3} \\
a_{1} b_{3}+a_{3} b_{1} \\
a_{2} b_{1}+a_{1} b_{2}
\end{array}\right)_{\mathbf{3_{2(2)}}},
$$

$$
\left(\begin{array}{l}
a_{1} \\
a_{2} \\
a_{3}
\end{array}\right)_{\mathbf{3_{1(1)}}} \otimes\left(\begin{array}{l}
b_{1} \\
b_{2} \\
b_{3}
\end{array}\right)_{\mathbf{3_{2(2)}}}=\left(a_{1} b_{1}+a_{2} b_{2}+a_{3} b_{3}\right)_{\mathbf{1_{2}}} \oplus\left(\begin{array}{l}
a_{1} b_{1}+\omega^{2} a_{2} b_{2}+\omega a_{3} b_{3} \\
-\omega a_{1} b_{1}-\omega^{2} a_{2} b_{2}-a_{3} b_{3}
\end{array}\right)_{\mathbf{2_{1}}}
$$

$$
\begin{aligned}
&\oplus\left(\begin{array}{l}
a_{1} b_{2}+\omega^{2} a_{2} b_{3}+\omega a_{3} b_{1} \\
-\omega a_{1} b_{3}-\omega^{2} a_{2} b_{1}-a_{3} b_{2}
\end{array}\right)_{\mathbf{2_{2}}} \\
&\oplus\left(\begin{array}{l}
a_{1} b_{3}+\omega^{2} a_{2} b_{1}+\omega a_{3} b_{2} \\
-\omega a_{1} b_{2}-\omega^{2} a_{2} b_{3}-a_{3} b_{1}
\end{array}\right)_{\mathbf{2_{3}}} \\
&\oplus\left(\begin{array}{l}
a_{1} b_{3}+a_{2} b_{1}+a_{3} b_{2} \\
-a_{1} b_{2}-a_{2} b_{3}-a_{3} b_{1}
\end{array}\right)_{\mathbf{2_{4}}},
\end{aligned}
$$

$$
\left(\begin{array}{l}
a_{1} \\
a_{2} \\
a_{3}
\end{array}\right)_{\mathbf{3_{1(2)}}} \otimes\left(\begin{array}{l}
b_{1} \\
b_{2} \\
b_{3}
\end{array}\right)_{\mathbf{3_{2(1)}}}=\left(a_{1} b_{1}+a_{2} b_{2}+a_{3} b_{3}\right)_{\mathbf{1_{2}}} \oplus\left(\begin{array}{l}
a_{1} b_{1}+\omega^{2} a_{2} b_{2}+\omega a_{3} b_{3} \\
-\omega a_{1} b_{1}-\omega^{2} a_{2} b_{2}-a_{3} b_{3}
\end{array}\right)_{\mathbf{2_{1}}}
$$

$$
\begin{aligned}
&\oplus\left(\begin{array}{l}
a_{1} b_{2}+\omega^{2} a_{2} b_{3}+\omega a_{3} b_{1} \\
-\omega a_{1} b_{3}-\omega^{2} a_{2} b_{1}-a_{3} b_{2}
\end{array}\right)_{\mathbf{2_{2}}} \\
&\oplus\left(\begin{array}{l}
a_{1} b_{3}+\omega^{2} a_{2} b_{1}+\omega a_{3} b_{2} \\
-\omega a_{1} b_{2}-\omega^{2} a_{2} b_{3}-a_{3} b_{1}
\end{array}\right)_{\mathbf{2_{3}}} \\
&\oplus\left(\begin{array}{l}
a_{1} b_{3}+a_{2} b_{1}+a_{3} b_{2} \\
-a_{1} b_{2}-a_{2} b_{3}-a_{3} b_{1}
\end{array}\right)_{\mathbf{2_{4}}},
\end{aligned}
$$

$$
\left(\begin{array}{l}
a_{1} \\
a_{2} \\
a_{3}
\end{array}\right)_{\mathbf{3_{1(2)}}} \otimes\left(\begin{array}{l}
b_{1} \\
b_{2} \\
b_{3}
\end{array}\right)_{\mathbf{3_{2(2)}}}=\left(\begin{array}{l}
a_{1} b_{1} \\
a_{2} b_{2} \\
a_{3} b_{3}
\end{array}\right)_{\mathbf{3_{2(1)}}} \oplus\left(\begin{array}{l}
a_{3} b_{2}-a_{2} b_{3} \\
a_{1} b_{3}-a_{3} b_{1} \\
a_{2} b_{1}-a_{1} b_{2}
\end{array}\right)_{\mathbf{3_{1(1)}}} \oplus\left(\begin{array}{l}
a_{3} b_{2}+a_{2} b_{3} \\
a_{1} b_{3}+a_{3} b_{1} \\
a_{2} b_{1}+a_{1} b_{2}
\end{array}\right)_{\mathbf{3_{2(1)}}},
$$

$$
\left(\begin{array}{l}
a_{1} \\
a_{2} \\
a_{3}
\end{array}\right)_{\mathbf{3_{2(1)}}} \otimes\left(\begin{array}{l}
b_{1} \\
b_{2} \\
b_{3}
\end{array}\right)_{\mathbf{3_{2(2)}}}=\left(a_{1} b_{1}+a_{2} b_{2}+a_{3} b_{3}\right)_{\mathbf{1_{1}}} \oplus\left(\begin{array}{l}
a_{1} b_{1}+\omega^{2} a_{2} b_{2}+\omega a_{3} b_{3} \\
\omega a_{1} b_{1}+\omega^{2} a_{2} b_{2}+a_{3} b_{3}
\end{array}\right)_{\mathbf{2_{1}}}
$$

$$
\begin{aligned}
&\oplus\left(\begin{array}{l}
a_{1} b_{2}+\omega^{2} a_{2} b_{3}+\omega a_{3} b_{1} \\
\omega a_{1} b_{3}+\omega^{2} a_{2} b_{1}+a_{3} b_{2}
\end{array}\right)_{\mathbf{2_{2}}} \\
&\oplus\left(\begin{array}{l}
a_{1} b_{3}+\omega^{2} a_{2} b_{1}+\omega a_{3} b_{2} \\
\omega a_{1} b_{2}+\omega^{2} a_{2} b_{3}+a_{3} b_{1}
\end{array}\right)_{\mathbf{2_{3}}} \\
&\oplus\left(\begin{array}{l}
a_{1} b_{3}+a_{2} b_{1}+a_{3} b_{2} \\
a_{1} b_{2}+a_{2} b_{3}+a_{3} b_{1}
\end{array}\right)_{\mathbf{2_{4}}},
\end{aligned}
$$

$$
\left(\begin{array}{l}
a_{1} \\
a_{2} \\
\end{array}\right)_{\mathbf{2_{s}}} \otimes\left(\begin{array}{l}
b_{1} \\
b_{2} \\
\end{array}\right)_{\mathbf{2_{s}}}=\left(
a_{1} b_{2}+a_{2} b_{1}\\
\right)_{\mathbf{1_{1}}} \oplus\left(
a_{1} b_{2}-a_{2} b_{1}\\
\right)_{\mathbf{1_{2}}} \oplus\left(\begin{array}{l}
a_{2} b_{2} \\
a_{1} b_{1} \\
\end{array}\right)_{\mathbf{2_{s}}} (s=1,2,3,4),
$$

$$
\left(\begin{array}{l}
a_{1} \\
a_{2} \\
\end{array}\right)_{\mathbf{2_{1}}} \otimes\left(\begin{array}{l}
b_{1} \\
b_{2} \\
\end{array}\right)_{\mathbf{2_{2}}}=\left(\begin{array}{l}
a_{2}b_{2} \\
a_{1}b_{1}\\
\end{array}\right)_{\mathbf{2_{3}}} \oplus\left(\begin{array}{l}
a_{2} b_{1} \\
a_{1} b_{2} \\
\end{array}\right)_{\mathbf{2_{4}}},
$$

$$
\left(\begin{array}{l}
a_{1} \\
a_{2} \\
\end{array}\right)_{\mathbf{2_{1}}} \otimes\left(\begin{array}{l}
b_{1} \\
b_{2} \\
\end{array}\right)_{\mathbf{2_{3}}}=\left(\begin{array}{l}
a_{2}b_{2} \\
a_{1}b_{1}\\
\end{array}\right)_{\mathbf{2_{2}}} \oplus\left(\begin{array}{l}
a_{2} b_{1} \\
a_{1} b_{2} \\
\end{array}\right)_{\mathbf{2_{4}}},
$$

$$
\left(\begin{array}{l}
a_{1} \\
a_{2} \\
\end{array}\right)_{\mathbf{2_{1}}} \otimes\left(\begin{array}{l}
b_{1} \\
b_{2} \\
\end{array}\right)_{\mathbf{2_{4}}}=\left(\begin{array}{l}
a_{1}b_{2} \\
a_{2}b_{1}\\
\end{array}\right)_{\mathbf{2_{2}}} \oplus\left(\begin{array}{l}
a_{1} b_{1} \\
a_{2} b_{2} \\
\end{array}\right)_{\mathbf{2_{3}}},
$$

$$
\left(\begin{array}{l}
a_{1} \\
a_{2} \\
\end{array}\right)_{\mathbf{2_{2}}} \otimes\left(\begin{array}{l}
b_{1} \\
b_{2} \\
\end{array}\right)_{\mathbf{2_{3}}}=\left(\begin{array}{l}
a_{2}b_{2} \\
a_{1}b_{1}\\
\end{array}\right)_{\mathbf{2_{1}}} \oplus\left(\begin{array}{l}
a_{1} b_{2} \\
a_{2} b_{1} \\
\end{array}\right)_{\mathbf{2_{4}}},
$$
 
$$
\left(\begin{array}{l}
a_{1} \\
a_{2} \\
\end{array}\right)_{\mathbf{2_{2}}} \otimes\left(\begin{array}{l}
b_{1} \\
b_{2} \\
\end{array}\right)_{\mathbf{2_{4}}}=\left(\begin{array}{l}
a_{1}b_{1} \\
a_{2}b_{2}\\
\end{array}\right)_{\mathbf{2_{1}}} \oplus\left(\begin{array}{l}
a_{1} b_{2} \\
a_{2} b_{1} \\
\end{array}\right)_{\mathbf{2_{3}}},
$$

$$
\left(\begin{array}{l}
a_{1} \\
a_{2} \\
\end{array}\right)_{\mathbf{2_{3}}} \otimes\left(\begin{array}{l}
b_{1} \\
b_{2} \\
\end{array}\right)_{\mathbf{2_{4}}}=\left(\begin{array}{l}
a_{1}b_{2} \\
a_{2}b_{1}\\
\end{array}\right)_{\mathbf{2_{1}}} \oplus\left(\begin{array}{l}
a_{1} b_{1} \\
a_{2} b_{2} \\
\end{array}\right)_{\mathbf{2_{2}}}.
$$ 
\end{appendices}


\begin{thebibliography}{100}

\bibitem{Altarelli:2010gt}
G.~Altarelli and F.~Feruglio,
Rev. Mod. Phys. \textbf{82}, 2701-2729 (2010).


\bibitem{Harrison:2002er}
P.~F.~Harrison, D.~H.~Perkins and W.~G.~Scott,
Phys. Lett. B \textbf{530}, 167 (2002).

\bibitem{Harrison:2002kp}
P.~F.~Harrison and W.~G.~Scott,
Phys. Lett. B \textbf{535}, 163-169 (2002).


\bibitem{Xing:2006xa}
Z.~Z.~Xing, H.~Zhang and S.~Zhou,
Phys. Lett. B \textbf{641}, 189-197 (2006).


\bibitem{He:2003rm}
X.~G.~He and A.~Zee,
Phys. Lett. B \textbf{560}, 87-90 (2003).




\bibitem{DoubleChooz:2011ymz}
Y.~Abe \textit{et al.} [Double Chooz],
Phys. Rev. Lett. \textbf{108}, 131801 (2012).

\bibitem{DayaBay:2012fng}
F.~P.~An \textit{et al.} [Daya Bay],
Phys. Rev. Lett. \textbf{108}, 171803 (2012).

\bibitem{DayaBay:2014fud}
F.~P.~An \textit{et al.} [Daya Bay],
Phys. Rev. D \textbf{90}, no.7, 071101 (2014).

\bibitem{RENO:2012mkc}
J.~K.~Ahn \textit{et al.} [RENO],
Phys. Rev. Lett. \textbf{108}, 191802 (2012).

\bibitem{T2K:2013ppw}
K.~Abe \textit{et al.} [T2K],
Phys. Rev. Lett. \textbf{112}, 061802 (2014).









\bibitem{Xing:2011at}
Z.~Z.~Xing,
Chin. Phys. C \textbf{36}, 101-105 (2012).

\bibitem{Zhou:2011nu}
S.~Zhou,
Phys. Lett. B \textbf{704}, 291-295 (2011).

\bibitem{Araki:2011wn}
T.~Araki,
Phys. Rev. D \textbf{84}, 037301 (2011).

\bibitem{Haba:2011nv}
N.~Haba and R.~Takahashi,
Phys. Lett. B \textbf{702}, 388-393 (2011).

\bibitem{Chao:2011sp}
W.~Chao and Y.~j.~Zheng,
JHEP \textbf{02}, 044 (2013).

\bibitem{Zhang:2011aw}
H.~Zhang and S.~Zhou,
Phys. Lett. B \textbf{704}, 296-302 (2011).

\bibitem{Rodejohann:2011uz}
W.~Rodejohann, H.~Zhang and S.~Zhou,
Nucl. Phys. B \textbf{855}, 592-607 (2012).

\bibitem{Marzocca:2011dh}
D.~Marzocca, S.~T.~Petcov, A.~Romanino and M.~Spinrath,
JHEP \textbf{11}, 009 (2011).

\bibitem{Antusch:2011ic}
S.~Antusch, S.~F.~King, C.~Luhn and M.~Spinrath,
Nucl. Phys. B \textbf{856}, 328-341 (2012).

\bibitem{Dev:2011bd}
S.~Dev, S.~Gupta and R.~Raman Gautam,
Phys. Lett. B \textbf{704}, 527-533 (2011).

\bibitem{Ge:2011qn}
S.~F.~Ge, D.~A.~Dicus and W.~W.~Repko,
Phys. Rev. Lett. \textbf{108}, 041801 (2012).

\bibitem{Ge:2011ih}
S.~F.~Ge, D.~A.~Dicus and W.~W.~Repko,
Phys. Lett. B \textbf{702}, 220-223 (2011).

\bibitem{Ludl:2011vv}
P.~O.~Ludl, S.~Morisi and E.~Peinado,
Nucl. Phys. B \textbf{857}, 411-423 (2012).

\bibitem{Joshipura:2011rr}
A.~S.~Joshipura and K.~M.~Patel,
JHEP \textbf{09}, 137 (2011).

\bibitem{Morisi:2011pm}
S.~Morisi, K.~M.~Patel and E.~Peinado,
Phys. Rev. D \textbf{84}, 053002 (2011).

\bibitem{BhupalDev:2011gi}
P.~S.~Bhupal Dev, R.~N.~Mohapatra and M.~Severson,
Phys. Rev. D \textbf{84}, 053005 (2011).

\bibitem{deAdelhartToorop:2011nfg}
R.~de Adelhart Toorop, F.~Feruglio and C.~Hagedorn,
Phys. Lett. B \textbf{703}, 447-451 (2011).

\bibitem{Adulpravitchai:2011rq}
A.~Adulpravitchai and R.~Takahashi,
JHEP \textbf{09}, 127 (2011).

\bibitem{Cao:2011cp}
Q.~H.~Cao, S.~Khalil, E.~Ma and H.~Okada,
Phys. Rev. D \textbf{84}, 071302 (2011).

\bibitem{Araki:2011qy}
T.~Araki and C.~Q.~Geng,
JHEP \textbf{09}, 139 (2011).

\bibitem{Rashed:2011xe}
A.~Rashed,
Nucl. Phys. B \textbf{874}, 679-697 (2013).

\bibitem{Rashed:2011zs}
A.~Rashed and A.~Datta,
Phys. Rev. D \textbf{85}, 035019 (2012).

\bibitem{Aranda:2011rt}
A.~Aranda, C.~Bonilla and A.~D.~Rojas,
Phys. Rev. D \textbf{85}, 036004 (2012).

\bibitem{Meloni:2011ac}
D.~Meloni,
JHEP \textbf{02}, 090 (2012).

\bibitem{King:2011ab}
S.~F.~King and C.~Luhn,
JHEP \textbf{03}, 036 (2012).



\bibitem{Haba:2006dz}
N.~Haba, A.~Watanabe and K.~Yoshioka,
Phys. Rev. Lett. \textbf{97}, 041601 (2006).

\bibitem{He:2006qd}
X.~G.~He and A.~Zee,
Phys. Lett. B \textbf{645}, 427-431 (2007).

\bibitem{Grimus:2008tt}
W.~Grimus and L.~Lavoura,
JHEP \textbf{09}, 106 (2008).

\bibitem{Ishimori:2010fs}
H.~Ishimori, Y.~Shimizu, M.~Tanimoto and A.~Watanabe,
Phys. Rev. D \textbf{83}, 033004 (2011).

\bibitem{Shimizu:2011xg}
Y.~Shimizu, M.~Tanimoto and A.~Watanabe,
Prog. Theor. Phys. \textbf{126}, 81-90 (2011).

\bibitem{He:2011gb}
X.~G.~He and A.~Zee,
Phys. Rev. D \textbf{84}, 053004 (2011).


\bibitem{deMedeirosVarzielas:2012cet}
I.~de Medeiros Varzielas and D.~Pidt,
JHEP \textbf{03}, 065 (2013).

\bibitem{Loualidi:2021qoq}
M.~A.~Loualidi,
[arXiv:2104.13734 [hep-ph]].

\bibitem{Zhao:2020cjm}
Z.~H.~Zhao, X.~Zhang, S.~S.~Jiang and C.~X.~Yue,
Int. J. Mod. Phys. A \textbf{35}, no.07, 2050039 (2020).

\bibitem{King:2019vhv}
S.~F.~King and Y.~L.~Zhou,
Phys. Rev. D \textbf{101}, no.1, 015001 (2020).

\bibitem{Novichkov:2018yse}
P.~P.~Novichkov, S.~T.~Petcov and M.~Tanimoto,
Phys. Lett. B \textbf{793}, 247-258 (2019).

\bibitem{Gautam:2018izb}
R.~R.~Gautam,
Phys. Rev. D \textbf{97}, no.5, 055022 (2018).

\bibitem{Rodejohann:2017lre}
W.~Rodejohann and X.~J.~Xu,
Phys. Rev. D \textbf{96}, no.5, 055039 (2017).

\bibitem{Luhn:2013lkn}
C.~Luhn,
Nucl. Phys. B \textbf{875}, 80-100 (2013).

\bibitem{King:2011zj}
S.~F.~King and C.~Luhn,
JHEP \textbf{09}, 042 (2011).

\bibitem{Kumar:2010qz}
S.~Kumar,
Phys. Rev. D \textbf{82}, 013010 (2010)
[erratum: Phys. Rev. D \textbf{85}, 079904 (2012)].


\bibitem{Dev:2022krz}
S.~Dev and D.~Raj,
Adv. High Energy Phys. \textbf{2022}, 4952562 (2022).

\bibitem{Grimus:2009xw}
W.~Grimus, L.~Lavoura and A.~Singraber,
Phys. Lett. B \textbf{686}, 141-145 (2010).

\bibitem{Albright:2008rp}
C.~H.~Albright and W.~Rodejohann,
Eur. Phys. J. C \textbf{62}, 599-608 (2009).

\bibitem{Zhao:2021efc}
Z.~h.~Zhao, X.~Y.~Zhao and H.~C.~Bao,
Phys. Rev. D \textbf{105}, no.3, 035011 (2022).

\bibitem{Ding:2020vud}
G.~J.~Ding, J.~N.~Lu and J.~W.~F.~Valle,
Phys. Lett. B \textbf{815}, 136122 (2021).



\bibitem{Lam:2006wy}
C.~S.~Lam,
Phys. Lett. B \textbf{640}, 260-262 (2006).

\bibitem{Harrison:2004he}
P.~F.~Harrison and W.~G.~Scott,
Phys. Lett. B \textbf{594}, 324-332 (2004).

\bibitem{Friedberg:2006it}
R.~Friedberg and T.~D.~Lee,
HEPNP \textbf{30}, 591-598 (2006).








\bibitem{Jarlskog:1985ht}
C.~Jarlskog,
Phys. Rev. Lett. \textbf{55}, 1039 (1985).

\bibitem{Bilenky:1987ty}
S.~M.~Bilenky and S.~T.~Petcov,
Rev. Mod. Phys. \textbf{59}, 671 (1987)
[erratum: Rev. Mod. Phys. \textbf{61}, 169 (1989); erratum: Rev. Mod. Phys. \textbf{60}, 575-575 (1988)].

\bibitem{Krastev:1988yu}
P.~I.~Krastev and S.~T.~Petcov,
Phys. Lett. B \textbf{205}, 84-92 (1988).


\bibitem{Barabash:2011row}
A.~S.~Barabash,
J. Phys. Conf. Ser. \textbf{375}, 042012 (2012).

\bibitem{KamLAND-Zen:2016pfg}
A.~Gando \textit{et al.} [KamLAND-Zen],
Phys. Rev. Lett. \textbf{117}, no.8, 082503 (2016).

\bibitem{NEXT:2009vsd}
F.~Granena \textit{et al.} [NEXT],
[arXiv:0907.4054 [hep-ex]].

\bibitem{NEXT:2013wsz}
J.~J.~Gomez-Cadenas \textit{et al.} [NEXT],
Adv. High Energy Phys. \textbf{2014}, 907067 (2014).

\bibitem{Licciardi:2017oqg}
C.~Licciardi [nEXO],
J. Phys. Conf. Ser. \textbf{888}, no.1, 012237 (2017).




\bibitem{Planck:2018vyg}
N.~Aghanim \textit{et al.} [Planck],
Astron. Astrophys. \textbf{641}, A6 (2020)
[erratum: Astron. Astrophys. \textbf{652}, C4 (2021)].
\bibitem{deSalas:2020pgw}
P.~F.~de Salas, D.~V.~Forero, S.~Gariazzo, P.~Mart\'\i{}nez-Mirav\'e, O.~Mena, C.~A.~Ternes, M.~T\'ortola and J.~W.~F.~Valle,
JHEP \textbf{02}, 071 (2021).

\bibitem{Ishimori:2010au}
H.~Ishimori, T.~Kobayashi, H.~Ohki, Y.~Shimizu, H.~Okada and M.~Tanimoto,
Prog. Theor. Phys. Suppl. \textbf{183}, 1-163 (2010).


\bibitem{Loualidi:2021qoq}
M.~A.~Loualidi,
[arXiv:2104.13734 [hep-ph]].
\bibitem{Altarelli:2005yp}
G.~Altarelli and F.~Feruglio,
Nucl. Phys. B \textbf{720}, 64-88 (2005).

\bibitem{King:2013eh}
S.~F.~King and C.~Luhn,
Rept. Prog. Phys. \textbf{76}, 056201 (2013).

\end{thebibliography}

\end{document}